\documentclass[twocolumn]{aastex631}

\usepackage{amsmath}

\usepackage{multirow}
\usepackage{hyperref}
\usepackage[noabbrev]{cleveref}  
\usepackage{float}  

\begin{document}

\title{Persistent radio sources associated with fast radio bursts: Implications from magnetar progenitors}

\author[0000-0003-1214-0521]{Sk. Minhajur Rahaman}
\affiliation{Astrophysics Research Center of the Open University (ARCO), The Open University of
Israel \\  P.O Box 808, Ra’anana 4353701, Israel}

\author{Sandeep Kumar Acharya}
\affiliation{Astrophysics Research Center of the Open University (ARCO), The Open University of
Israel \\  P.O Box 808, Ra’anana 4353701, Israel}

\author[0000-0001-7833-1043]{Paz Beniamini}
\affiliation{Astrophysics Research Center of the Open University (ARCO), The Open University of
Israel \\  P.O Box 808, Ra’anana 4353701, Israel}
\affiliation{Department of Natural Sciences, The Open University of Israel \\ P.O Box 808, Ra’anana
4353701, Israel}
\affiliation{Department of Physics, The George Washington University\\ 725 21st Street NW, Washing-
ton, DC 20052, USA}

\author{Jonathan Granot}
\affiliation{Astrophysics Research Center of the Open University (ARCO), The Open University of
Israel \\  P.O Box 808, Ra’anana 4353701, Israel}
\affiliation{Department of Natural Sciences, The Open University of Israel \\ P.O Box 808, Ra’anana
4353701, Israel}
\affiliation{Department of Physics, The George Washington University\\ 725 21st Street NW, Washing-
ton, DC 20052, USA}




\begin{abstract}
The rare association of three persistent radio sources (confirmed PRS1 and PRS2, candidate PRS3) with repeating fast radio bursts (FRB 20121102A, 20190520B, 20201124A) offers a unique probe into their magneto-ionic environments. PRSs are attributed to synchrotron emission from relativistic charged particles of magnetar wind nebula (MWN), powered by spin-down magnetohydrodynamic wind or internal magnetic field decay. Using a multizone hydrodynamic model, we track MWN evolution to constrain magnetar progenitor properties. For PRS1 and PRS2, we find an equipartition radius $R_{\rm eq} \sim 0.1 \;$ pc that is consistent with the radio scintillation estimates ($> 0.03$ pc) and radio imaging limits ($<0.7$ pc). This compact size favors low expansion speeds and large initial spin periods, $P_{\rm i}\gtrsim10\;$ms, ruling out millisecond magnetar progenitors. Given $P_{\rm i}\gtrsim10\;$ ms, a current size of $\sim 0.1$ pc, a supernova kinetic energy $E_{\rm SN} \sim 10^{50}-10^{51} \;$ erg and an ejecta mass $M \sim 3-10 \; M_{\odot}$, the PRS age is $t \sim 10-10^{2}$ yr. PRSs with $t>20$ years require an internal field ($B_{\rm int}\sim10^{16}-10^{16.5}\;$G) with a decay timescale $t_{\rm d}\sim10-10^{2.5}$ yr. The slowest field decay ($t_\mathrm{d,max}\sim 500$ yr) favors sub-energetic supernovae ($E_{\rm SN}\sim10^{50}\;$erg) with massive ejecta ($M\gtrsim 10\;M_{\odot}$) and low ionization fraction ($\sim3\%$). For the sub-energetic scenario, for the confirmed PRSs we predict a cooling break at 100–150 GHz at 20–40 $\mu\text{Jy}$ and self-absorption near 200 MHz at 180 $\mu$Jy. For PRS3, a rotation-powered MWN is viable only if $t \sim 10$ yr; an inverted spectrum beyond 150 GHz would rule out this scenario.
\end{abstract}

\keywords{ Radio transient sources (2008)  --- Radio bursts (1339) --- Radio continuum emission (1340) --- Neutron stars (1108) }


\section{Introduction} \label{sec:intro}
Fast radio bursts (FRBs) are highly coherent radio transients, typically of millisecond duration, originating from extragalactic distances. Until now, thousands of FRBs have been discovered \citep{Lorimer2007,Petroff2022,CHIME2021}. FRBs are broadly categorized into two types: repeaters that repeat their activity on widely variable timescales, and non-repeaters that appear
to be one-time events \citep{Chime2023}. Observationally, repeaters are rarer compared to non-repeaters. Only a small fraction of repeaters and nonrepeaters have been localized to their host galaxies \citep{Chatterjee2017,Zhang2023,Connor2024}. Despite significant progress in their detection and localization, the origins of FRBs and their immediate magneto-ionic environments remain poorly understood \citep{Platts2019,Zhang2023}.

Persistent radio sources (hereafter PRSs) are compact sub-parsec ($< 1\;$pc), highly luminous 
($L_{\nu} > 10^{29}$ ${\rm erg\; s^{-1}\; Hz^{-1}}$) 
quasi-steady continuum radio sources at GHz frequencies, found in the off-nuclear regions of host dwarf galaxies, with luminosities unrelated to star formation activity \citep{Law22,Ibik24}.
The association of PRSs with active repeating FRBs offers critical insights into FRB progenitors and their local environments. The PRSs linked to FRB 20121102A and FRB 20190520B exhibit similar luminosities and negative spectral indices. Both sources show variable rotation measures (RM), indicating turbulent environments near their progenitors \citep{Chatterjee2017,Niu2022,Michilli2018,Anna-Thomas2023}. A third candidate PRS, associated with FRB 20201124A, is less luminous and features an inverted spectrum \citep{Bruni2023}. Notably, PRS luminosities correlate with their RMs, suggesting a relationship between nebular properties and the magneto-ionic environments. Section \S \ref{desc_sources} provides a detailed discussion of these sources. 

Magnetars are leading candidates \citep{popov_2010,katz16,Kumar+17,wadiasingh2019,2021Univ....7...56L,BK2023,2024arXiv241019043B} for FRB progenitors, supported by the detection of a low-luminosity FRB from the galactic magnetar SGR 1935+2154 \citep{Bochenek2020,2020Natur.587...54C,MBSM2020,LKZ2020}. The link between PRSs and FRBs provides a valuable framework to test the magnetar progenitor hypothesis. In this interpretation, the PRS is thought to arise from synchrotron emission produced by energetic charged particles (relativistic electrons and positrons) within the magnetar wind nebulae (MWN). However, the power source for MWN remains uncertain -- it may originate from rotational energy through magnetic dipole braking 
or from the decay of the internal magnetic field.

For example, studies of the PRS associated with FRB 20121102A suggest conflicting origins. Some analyses indicate the emission could be rotation-powered \citep{Kashiyama17,2024arXiv241219358B}, while others propose internal magnetic field decay as the dominant driver \citep{Beloborodov2017,MM2018}. Evidence increasingly supports the presence of stronger internal magnetic fields in magnetars \citep{1996ApJ...473..322T,2012MNRAS.422.2878D,2015RPPh...78k6901T,Granot2017,2024MNRAS.532.4535M}. Specifically, \citet{Granot2017} (hereafter G17) demonstrated that \textit{Swift} J1834-0846, the only directly observed MWN around a known Galactic magnetar, can only be powered by internal magnetic field decay, reinforcing the plausibility of magnetic powering of PRSs. 

However, many open issues still need 
to be addressed in the MWN interpretation of PRSs, and several key issues remain unresolved. 
In previously published magnetically powered PRS models, the assumed internal magnetic field decay timescale ($ \sim\!0.2\,$–$\,0.6\;$yr) is at least four orders of magnitude shorter than the decay timescale $ t_{\rm d}\gtrsim 10^3$ years, estimated for Galactic magnetars \citep[][where $\sim1\;$kyr and $\sim10\;$kyr were estimated for the dipole and internal fields, respectively]{2012MNRAS.422.2878D}.

Additionally, \cite{2024arXiv241219358B} argue that the magnetic decay and rotation-powered scenarios are nearly indistinguishable and largely independent of the supernova progenitor. The nature of the progenitor itself is also debated, with \cite{2024arXiv241219358B} favoring an ultra-stripped supernova, while while \cite{Murase16,Metzgar17} propose a superluminous supernova. Both models, however, suggest a millisecond spin period for the progenitor magnetar. Moreover, earlier studies did not consider the potential role of enhanced supernova remnant expansion in selecting or favoring a specific type of magnetar progenitor. Additionally, they assumed a non-decaying magnetic energy content within the nebula, which may lead to overestimates of the long-term magnetic field strength. We address and expand upon these limitations in \S ~\ref{rob_conc}. As we will demonstrate, consideration of adiabatic losses has important implications for favoring a particular magnetar progenitor (see \S \ref{conc} later for an expanded discussion).    

Furthermore, earlier studies of PRSs associated with FRBs lacked access to many observational constraints that became available later (see \S \ref{rob_conc}), leading to broad age estimates ranging from a decade to a thousand years. Besides, a crucial gap remains in the lack of predicted observable signatures that could definitively distinguish between competing scenarios.

A comprehensive investigation of the magnetar progenitor parameter space, accounting for both rotational and magnetically powered scenarios, is essential to satisfy the constraints imposed by FRBs and their associated PRSs. This study seeks to determine whether either scenario can be definitively ruled out for powering the known sources. The objective is to predict the self-absorption and cooling frequencies in the PRS spectrum, as well as the age of the nebula. These predictions can help differentiate between power sources or potentially exclude the MWN interpretation entirely.

The structure of this paper is as follows: \S\,\ref{desc_sources} provides a detailed discussion of the PRSs associated with the three FRBs. \S\,\ref{setup} outlines the evolution of MWN's energy and radius under combined rotational and magnetic decay power. \S\,\ref{MWN_param} constrains the magnetar parameter space for MWNe and examines the implications of these constraints. \S\, \ref{rob_conc} compares our findings with previous findings in the literature. \S\,\ref{conc} summarizes our key results and suggests a future observational campaign.

\section{Observations} \label{desc_sources}

This section presents the key observational findings for the three FRB-PRS systems. A summary of their properties is provided in Table \ref{tab:properties}, and a brief description of each system is given below:  

\begin{itemize}  
    \item \textbf{PRS associated with FRB 20121102A}: This source is the most well-studied and well-monitored PRS. It is located in a dwarf galaxy with a positional offset in the off-nuclear region. Its radio spectrum has been monitored from 500 MHz (uGMRT) to 22 GHz (VLA). While earlier studies \citep{Chatterjee2017} indicated a cooling break at 10 GHz, later studies \citep{2023ApJ...958..185C} have shown that this identification is spurious due to the effects of scintillation. The same study suggests a robust lower limit on the source size ($>0.03\;$pc). Radio imaging studies such as \cite{Marcote2017} imply an upper limit on the PRS size of $<0.7\;$pc. Radio scintillation studies \citep{2023ApJ...958..185C} point to an even compacter size $\sim\!0.2\;$pc (this is consistent with results from equipartition analysis in \S \ref{spec_radio}).
    
    \item \textbf{PRS associated with FRB 20190520B}: Its SED has been observed in the $\sim\,$1\,--\,6\;GHz range \citep{Niu2022}, showing a similarly flat profile, while its projected size is poorly constrained. The angular separation between the FRB and PRS is $<20\;$mas \citep{Bhandari2023}.  Moreover, its RM fluctuates significantly, showing sign reversals within a few months \citep{Anna-Thomas2023}.  

    \item \textbf{PRS associated with FRB 20201124A}: This third candidate PRS  
     has been reported \citep{Lanman2022,Bruni2023}. It has an inverted spectrum, albeit with large uncertainties. The PRS association with the FRB remains tentative due to a positional offset (angular separation $\sim $ 56 mas) between the PRS and FRB locations. \cite{Bruni2023} identified a potential correlation between the luminosities and rotation measures of this PRS, supporting a nebular origin.  
\end{itemize}

The third tentative source differs significantly from the first two. It has an inverted spectrum, a luminosity two orders of magnitude lower, and a projected size upper limit ($\sim$kpc), which is two orders of magnitude larger. Its rotation measure is much smaller, and the positional offset between the PRS and the repeater FRB is much larger ($\sim$ tens of pc). These differences make it a candidate rather than a confirmed association. 

The next section lays the groundwork for a comprehensive discussion of the observed properties and their implications for the nature of PRSs.

\begin{table*}
\footnotesize
\caption{Summary of the observed properties of the reported PRS associated with FRBs. Here, $z$ denotes the redshift of the source, DM represents the dispersion measure, RM stands for the rotation measure, $\nu$ is the observational frequency, $R_\mathrm{proj}$ is the projected radius of the PRS, $F_{\nu}$ is the observed flux density of the PRS, $\alpha$ is the spectral index (such that the flux density scales as $F_{\nu} \propto \nu^{\alpha}$), and $L_{\nu}$ is the spectral luminosity of the PRS. The references for the properties are numbered as, [1]  \href{https://ui.adsabs.harvard.edu/abs/2017ApJ...834L...7T}{\citep{Tendulkar17}}
[2] \href{https://doi.org/10.1038/nature17168}{\citep{Spitler2016}}, 
[3] \href{https://doi.org/10.1126/science.abn8837}{\citep{Niu2022}}, 
[4] \href{https://doi.org/10.3847/1538-4357/ac82d8}{\citep{Lanman2022}}, 
[5] \href{https://doi.org/10.3847/2041-8213/ac3aed}{\citep{Hilmarsson2021}}, 
[6] \href{https://doi.org/10.1038/nature20797}{\citep{Chatterjee2017}}, 
[7] \href{https://doi.org/10.3847/2041-8213/aa7b30}{\citep{Marcote2017}}, 
[8] \href{https://doi.org/10.3847/1538-4357/aca8c7}{\citep{Anna-Thomas2023}}, 
[9] \href{https://doi.org/10.1093/mnras/stad120}{\citep{Bhandari2023}}, 
[10] \href{https://doi.org/10.1093/mnras/stad785}{\citep{Bruni2023}}.
}
  \begin{center}
   \begin{tabular}{|l|l|c|c|c|} \hline 
   \multicolumn{2}{|c|}{Property}  & FRB20121102A  & FRB20190520B & FRB20201124A \\ \hline\hline 
     \multirow{3}{*}{\textbf{FRB}} 
     &
    $z$ (redshift) & 0.193 [\href{https://ui.adsabs.harvard.edu/abs/2017ApJ...834L...7T}{1}] & 0.241 [\href{https://doi.org/10.1126/science.abn8837}{3}] & 0.098 [\href{https://doi.org/10.3847/1538-4357/ac82d8}{4}] \\ \cline{2-5}
    & DM [pc cm$^{-3}]$ & 558 [\href{https://doi.org/10.1038/nature17168}{2}] & 1204 [\href{https://doi.org/10.1126/science.abn8837}{3}] & 413 [\href{https://doi.org/10.3847/1538-4357/ac82d8}{4}] \\ \cline{2-5}
    & RM [rad m$^{-2}$] & $1.4\times 10^5$ [\href{https://doi.org/10.3847/2041-8213/ac3aed}{5}] & [$-2.4\times 10^4-1.3\times 10^4]$ [\href{https://doi.org/10.3847/1538-4357/aca8c7}{8}] & 900 [\href{https://doi.org/10.1093/mnras/stad785}{10}] \\ \hline\hline
    \multirow{5}{*}{\textbf{PRS}} & $\nu$ [GHz] & 1\,--\,26 [\href{https://doi.org/10.1038/nature20797}{6}] & 1.5,\,3,\,5.5 [\href{https://doi.org/10.1126/science.abn8837}{3}] & 6,\,15,\,22 [\href{https://doi.org/10.1093/mnras/stad785}{10}]  \\ \cline{2-5}
     & $R_{\rm proj}$ [pc] & $<$ 0.7  (at 5 GHz) [\href{https://doi.org/10.3847/2041-8213/aa7b30}{7}] & $<$ 9  [\href{https://doi.org/10.1093/mnras/stad120}{9}] & $<$ 700  [\href{https://doi.org/10.1093/mnras/stad785}{10}] \\ \cline{2-5}
      & $F_{\nu}$ [$\mu$Jy] & 180 (3 GHz) [\href{https://doi.org/10.1038/nature20797}{6}] & 202 (3 GHz) [\href{https://doi.org/10.1126/science.abn8837}{3}] & 8 (6), 20 (15) and 30 (22 GHz) [\href{https://doi.org/10.1093/mnras/stad785}{9}]\\ \cline{2-5}
     & $\alpha$ (spectral index) & $-$0.2 ($<10$ GHz), $-$1 ($>10$ GHz) [\href{https://doi.org/10.1038/nature20797}{6}] & $-0.41\pm 0.04$ [\href{https://doi.org/10.1126/science.abn8837}{3}] & $1\pm 0.43$ [\href{https://doi.org/10.1093/mnras/stad785}{10}] \\ \cline{2-5}
     & $L_{\nu}$ [${\rm erg\;s^{-1}\;Hz^{-1}}$] & $\sim 2 \times 10^{29}$ (2 GHz) & $\sim 3 \times 10^{29}$ (2 GHz) & $\sim 2\times 10^{27}$ (6 GHz) \\ \hline 
    \end{tabular}
    \end{center}
\label{tab:properties}
\end{table*}

\section{Basic set-up}\label{setup}

The objective of the present section is to describe the environment around a magnetar and the effect it can have on the observables. We assume a spherically symmetric multizone-zone model in which the physical properties within any zone remain homogeneous.  

\begin{figure}
\centering
\begin{tabular}{c}
\includegraphics[scale=0.52]{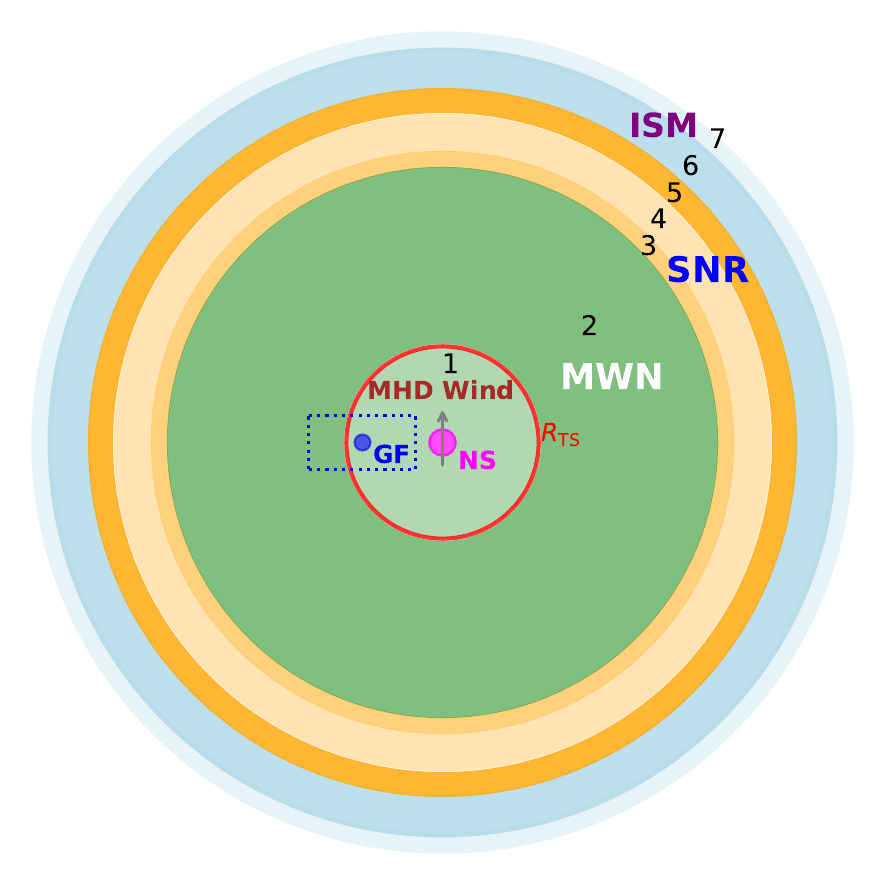}
\end{tabular}
\caption{A schematic (not to scale) depicting the environment surrounding a young magnetar (shown in pink) enclosed within a supernova remnant (SNR).  The accompanying table \ref{MWN_regions_table} gives the description of the various regions  and thjeir contribution to the observables (see text for a detailed explanation).  }  \label{MWN-GF_setup}
\end{figure}

\begin{deluxetable*}{rll}
\footnotesize
\tablewidth{0pt}
\tablecaption{Description of regions in Fig.~\ref{MWN-GF_setup} and their contributions to the observables \label{MWN_regions_table}}
\tablehead{
\colhead{Region} & \colhead{Description} & \colhead{Contribution}
}
\startdata
7 & Unshocked ISM (light blue) & -- \\  
6 & ISM shocked by the forward shock (FS) from the SNR-ISM interaction (dark blue) & -- \\  
5 & SNR ejecta shocked by the reverse shock (RS) from the SNR-ISM interaction (dark orange) & DM$_\mathrm{SNR}$ \\  
4 & Unshocked SNR ejecta (lighter orange) & DM$_\mathrm{SNR}$ \\  
3 & SNR ejecta shocked by the FS from the MW-SNR interaction & DM$_\mathrm{SNR}$ \\  
2 & MWN shocked by the RS from the MW-SNR interaction (dark green) & PRS \\  
1 & Cold MW carrying the majority of $L_\mathrm{sd}$ & -- \\   
$R_\mathrm{TS}$ & Termination shock radius, balancing hot MWN and cold MW & -- \\
\enddata
\tablecomments{The color references in parentheses correspond to the visualization in Fig.~\ref{MWN-GF_setup}. DM$_\mathrm{SNR}$ indicates contributions to dispersion measure from supernova remnant material.}
\end{deluxetable*}

Figure \ref{MWN-GF_setup} illustrates a schematic depiction of the environment surrounding a young magnetar system. The supernova remnant of mass $M_\mathrm{SNR}$ expanding at a velocity $v_\mathrm{SNR}$ carrying kinetic energy $E_\mathrm{SNR}$ (hereafter referred to as the SNR ejecta), confines an electron-positron magnetohydrodynamic (MHD) wind launched by the central magnetar. 
Given the compact nature of the PRS, we assume that its age is smaller than the Sedov-Taylor time, $t_\mathrm{dec}$, of the SNR ejecta. For times $t < t_\mathrm{dec}$, the SNR ejecta remains in the free-expansion phase, during which its radius grows linearly with time. The ejecta's coasting velocity, $v_\mathrm{SNR}$, can be expressed as:
\begin{equation}
    v_\mathrm{SNR} \approx 5.8 \times 10^{3} \; M_{\rm 3}^{-\frac{1}{2}} \; E_{\rm SNR,51}^{\frac{1}{2}}\;  \text{km s$^{-1}$} \;,
\end{equation}
where $M_{3} = M_{\rm SNR}/(3 M_{\odot})$ is the mass of the ejecta and $E_\mathrm{SNR,51} = E_{\rm SNR}/(10^{51} \;$erg) the total kinetic energy 
of the ejecta; $E_{\rm SNR}$ includes the mechanical energy from the supernova explosion, $E_\mathrm{SN}$, and the rotational energy, $E_\mathrm{rot}$, lost during the spin-down phase.\footnote{In principle, $E_\mathrm{SNR}$ should also account for the magnetically power outflows ejected during the field decay timescale, $t_\mathrm{d}$, if $t_\mathrm{d} < t_\mathrm{dec}$. However, for the parameter space explored here, the decay timescale $t_{\rm d}$ is consistently longer than the spin-down timescale $t_{\rm sd}$ and the magnetic energy smaller than $\max(E_{\rm rot},E_{\rm SN})$, allowing us to neglect the contribution of internal magnetic field decay to $v_\mathrm{SNR}$.} During the free-coasting phase, the nebula's radius can be assumed to match the inner radius of the SNR shell and is given by:
\begin{equation}
    R_\mathrm{n} = R_\mathrm{SNR} \approx 1.8 \times 10^{17} \; M_{\rm 3}^{-\frac{1}{2}} \; E_{\rm 51}^{\frac{1}{2}} \; t_{10}  \; \text{cm}   \hspace{0.2cm} \text{for $t <  t_\mathrm{dec}$}\;,  \label{neb_rad}
\end{equation}
where $t_{10} = t / 10 \; \mathrm{yr}$. The Sedov-Taylor time $t_\mathrm{dec}$, beyond which the free-coasting phase ends, is,
\begin{equation}
    t_\mathrm{dec} \approx 517 \; M_{\rm 3}^{\frac{5}{6}} \; E_{\rm SNR,51}^{-\frac{1}{2}} \; n_{\rm o}^{-\frac{1}{3}} \; \text{yr}\;,   \label{time_ST}
\end{equation}
where $n_{\rm o} = n/ 1\;  \text{cm$^{-3}$}$ is the particle density of the ambient medium into which the SNR is expanding.  

The MWN’s termination shock, located at radius $R_{\rm TS}$, marks the boundary where the ram pressure of the hot MWN balances that of the cold MHD wind. The ratio of the termination shock radius to the supernova remnant radius $R_{\rm TS}/R_{\rm SNR}$ can be expressed in the free-coasting phase as (see appendix \ref{app:term-rad})
\begin{equation}
    \frac{R_{\rm TS}}{R_{\rm SNR}} = \sqrt{\frac{v_\mathrm{SNR}}{c}} \approx 0.14 \; E_{\rm 51}^{\frac{1}{4}} \; M_{3}^{-\frac{1}{4}} \hspace{0.2cm} \text{for $ t \leq t_{\rm dec}$}\;. \label{RTS}
\end{equation}
Equation \ref{RTS} shows that the volume enclosed within the termination shock radius is minuscule compared to the total volume enclosed by the supernova remnant.

Interactions between the different media give rise to two shock fronts: a forward shock (FS) propagating outward into the slower material and a reverse shock (RS) propagating inward into the faster material. The regions in the schematic Fig.~\ref{MWN-GF_setup} are numbered 1 to 7.
Following \cite{Piro2016}, we attribute some of the DM contribution near the progenitor to the SNR ejecta. Regions 3 and 5, being shocked, contribute to the ionization of the ejecta, while Region 4, though unshocked, may also become ionized by ionizing photons. The combined DM contribution of the SNR ejecta (Regions 3, 4, and 5) is characterized by average ionization fraction $f_\mathrm{ion}$. The observed DM is the cumulative contribution from various sources along the propagation path of the FRB signal. These contributions include the Intergalactic Medium (IGM), the Interstellar Medium (ISM) of our Galaxy, the Galactic Halo, as well as the source and host contributions, which are redshifted by a factor of $1 + z$. The observed DM can therefore be expressed as:

\begin{equation}
{\rm DM} = {\rm DM_{IGM}} + {\rm DM_{ISM}} + {\rm DM_{halo}} + \frac{{\rm DM_{\rm src}} + {\rm DM_{host}}}{1 + z}
\end{equation}

Here, ${\rm DM_{IGM}}$ represents the contribution from the IGM, which arises from the large-scale structure of the universe. ${\rm DM_{ISM}}$ and ${\rm DM_{halo}}$ are contributions from our Milky Way's ISM and the surrounding Galactic Halo, respectively. For the analysis presented here, the $\mbox{DM}_\mathrm{src}$ is due to the MWN and the Supernova Remnant (SNR) ejecta. This term is denoted as ${\rm DM_{SNR}}$. The host term arises from the host galaxy at redshift $z$, which contains the source, and may also include contributions from its circumgalactic medium. The IGM contribution to DM from $z$ of 0.1 and 0.2 is around 90 and 180 pc cm$^{-3}$ respectively (e.g. \cite{Connor2024}). The host galaxy ISM also typically contributes $\lesssim 60\;\mbox{pc cm}^{-3}$ (e.g. \citealt{Acharya25}). Thus, most of the DM after subtracting the contribution from the IGM and the Milky way seems to be from the host galaxy and likely even the immediate environment around the FRB progenitor. 

Furthermore, RM can only arise from unpaired electrons - a MWN consisting purely of electron-positron plasma will not contribute to the RM. For RM, we consider that the contribution primarily comes from region $2'$ (not shown), which is embedded within the MWN, and contains both ions and pair plasma.
It is important to emphasize that the rotation measure (RM) is highly sensitive to the local environment and exhibits significant temporal variations. Our one-zone model does not account for this time variability, particularly the effects of turbulence, which can be crucial in young, energized nebulae. As a result, we do not impose constraints based on RM measurements in this work. Instead, we refer to some fiducial estimates in subsection \ref{RM_DM_correlation}.

Table \ref{MWN_regions_table} summarizes the contributions of the various regions to observable quantities depicted in Figure \ref{MWN-GF_setup}. Table \ref{tab:glossary} gives a comprehensive list of symbols used throughout the text and their meaning.

\subsection{Requirements for powering FRBs }\label{req_FRB}

Recent studies indicate that the energetics and high brightness temperatures associated with fast radio bursts (FRBs) suggest a very strong internal magnetic field  ($B_{\rm int}\geq 10^{15}$ G) at their source \citep{2018MNRAS.477.2470L,2024arXiv241019043B}. In a magnetar-based model for FRBs, the luminosity is thought to be driven by the decay of the internal magnetic field $B_\mathrm{int}$. However, $B_\mathrm{int}$ cannot be arbitrarily large. Numerical studies \citep{2009MNRAS.397..763B} indicate  $B_\mathrm{int} \leq 3 \times 10^{16}\;$G beyond which the magnetic field energy density would overcome the crustal binding energy. Moreover, the same numerical studies indicate that the ratio of the dipolar magnetic field strength to the internal magnetic field cannot be less than 1:30,
otherwise the magnetic field configurations become unstable. This sets a lower limit on the dipolar magnetic field strength at \( B_\mathrm{d} = 10^{15}\;\)G  when \( B_\mathrm{int} = 3 \times 10^{16}\;\)G. Following \citealt{2018MNRAS.477.2470L} we consider a fiducial dipolar magnetic field of at least $10^{14}\;$G (and correspondingly $B_{\rm int} = 10^{15}\;$G) to power an FRB, 
\begin{equation}
    B_{\rm d, min, FRB} \gtrsim 10^{14} \; \text{G}\;. \label{Bd_frb}
\end{equation}

\subsection{Energy injection}\label{eng_inj}

We consider here the two channels by which magnetars can power a MWN. The first channel is the spindown-powered relativistic MHD pair wind,
while the other is decay of the internal magnetic field that 
powers episodic \textit{electron-ion outflows} the most dramatic of which are magnetar giant flares. If the time-between giant-flare ejections is smaller than the system's age, the rate at which the magnetic energy is released can be approximated as continuous. An important difference between the spindown and magnetically powered phases is that in the former the charged particles are (almost) purely electron-positron pairs while for the latter the composition is (predominantly) baryonic electron-ion.   

We consider a magnetar with radius $R_\mathrm{NS}$ and moment of inertia $I$ with equatorial surface magnetic dipole field $B_d$ and angular frequency $\Omega$. The spin-down power $L_\mathrm{sd}$ is given by,
\begin{equation}
    L_{\rm sd} (t) = L_{0} \left( 1 + \frac{t}{t_{\rm sd}} \right)^{-2} \hspace{0.1cm} \text{for magnetic dipole breaking}. 
\end{equation}
The initial rotational  energy $E_{\rm rot}$, the initial spin-down luminosity $L_0$ and the spin-down time $t_{\rm sd}$ are approximated as, 
\begin{equation}
\begin{split}
&\ E_{\rm rot}= \frac{1}{2} I \Omega_{\rm i}^2  \approx 2\times 10^{52} \; P_{\rm i,-3}^{-2}\;  \text{erg}\;, \\
&\ L_\mathrm{0} = f\frac{B_{\rm d}^2 R_{\rm NS}^6\Omega_{\rm i}^4}{\rm c^3} \approx\; 5.8 \times 10^{47} P_{\rm i,-3}^{-4} \; f \;  B_{\rm d,14}^{2} \;  \text{erg s$^{-1}$}\;, \\ 
&\ t_\mathrm{sd}=\frac{I{\rm c}^3}{2f\Omega_i^2R_{\rm NS}^6B_\mathrm{d}^2} \approx 3\times 10^4 \; P_{\rm i,-3}^2 \; f^{-1}\;  B_{\rm d,14}^{-2} \; \text{s}\;, 
\end{split}\label{rot_parm}
\end{equation} 
where we used the moment of inertia $I=10^{45}$ g cm$^2$, neutron star radius $R_{\rm NS}=10^6$ cm, initial spin period $P_{i,-3}={P_{\rm i} }/{\rm 1\; ms}$ and the surface dipolar magnetic field strength $B_{\rm d}=10^{14}B_{\rm d,14}\;$G. Here the quantity $f$ is $\mathcal{O}(1)$. In the force-free magnetosphere limit $f=1+\sin^2\theta_B$ such that $1\leq f\leq2$. For simplicity we assume $f=1$ in our analysis. During $t_{\rm sd}$,  almost all of the initial rotational kinetic energy $L_{0} t_{\rm sd} \approx E_{\rm rot}$ is lost.

Following \cite{2024arXiv241116846B} the evolution of the internal magnetic field of a magnetar is given by,
\begin{equation}
    E_\mathrm{B}(t)=E_\mathrm{B,0}\left(1+\frac{\alpha_{\rm B} t}{t_\mathrm{d}}\right)^{-\frac{2}{\alpha_{\rm B} } }\;,
\end{equation}
which results in an average magnetic luminosity,
\begin{equation}
    L_\mathrm{B}(t)= L_{\rm B,0} \left( 1 + \frac{\alpha_{\rm B} t}{t_{\rm d}}\right)^{-\frac{2}{\alpha_{\rm B}} - 1}\;,   
\end{equation}
where the initial internal magnetic field energy $E_{\rm B,0}$,  the initial internal magnetic field decay timescale $t_{\rm d}$ and the initial magnetic field luminosity $L_{\rm B,0}$ can be represented as 
\begin{equation}
    \begin{split}
        &\ E_{\rm B,0}= \frac{1}{6} B_{\rm int}^2 R_{\rm NS}^3 \;  \approx  2\times 10^{49} \; B_{\rm int,16}^2  \; \text{erg}\;, \\ 
        &\   t_{\rm d}   \approx 10^{3} \; t_{\rm d,3} \,B_{\rm int,16}^{-\alpha_{\rm B}}\; \text{yr} \hspace{0.2cm} \text{for $\alpha_{\rm B} = 1$}\;, \\      
        &\ L_{\rm B,0} = \frac{\alpha_{\rm B} E_{\rm B,0}}{ t_{\rm d}} \approx 5.3 \times 10^{38} \; t_{\rm d,3}^{-1}\; B_{\rm int,16}^{2+ \alpha_{\rm B}} \; \text{erg s$^{-1}$} \;     \hspace{0.1cm} \text{for $\alpha_{\rm B} = 1$}\;,
    \end{split} \label{mag_param}
\end{equation}
where we assume that the fiducial internal decay timescale $t_{\rm d} \propto B_{\rm int}^{-\alpha_{\rm B}}$, such that $t_{\rm d,3} = t_{\rm d}/(10^{3} \; \text{yr})$ , which is an order of magnitude smaller than found by \citep{2012MNRAS.422.2878D}  for galactic magnetars ($t_{\rm d} = 10^{4}$ year for $B_\mathrm{int,16}$ =1). In contrast all the earlier magnetic decay-power models assume $t_\mathrm{d} \sim 0.2$ years (see \S \ref{rob_conc} ).  For the rest of the analysis we will assume $\alpha_{\rm B} = 1$. 

Eq.~(\ref{mag_param}) shows that the decay timescale has an inverse relation with the internal magnetic field strength. Since a strong internal magnetic field is a necessary requirement for powering FRBs (see \S \ref{req_FRB}), the corresponding internal magnetic field decay timescale is shorter.  
For a short decay timescale the injected energy in the nebula is degraded more, and vice versa. Thus, adiabatic loss has a prominent role to play in determining the energy content of the nebula as a function of time.

We assume that the non-thermal electrons carry a constant fraction $\epsilon_{\rm e}$  of the injected energy in the nebula. The instantaneous power $L_{\rm e}$ of the non-thermal electrons can be represented as 
\begin{equation}
    L_{\rm e} = \epsilon_{\rm e} L_{\rm inj} =
    \begin{cases}
    \;\epsilon_{\rm e} L_{\rm sd} \hspace{0.7cm} \text{for spindown power},        \\
    \;\epsilon_{\rm e} \epsilon_{\rm out} L_{\rm B} \hspace{0.2cm} \text{for magnetic decay power}.
    \end{cases} \label{part_lum}
\end{equation}

Eq.~(\ref{part_lum}) shows that
in general, only a fraction $\epsilon_{\rm out}$ of the magnetic power $L_{\rm B}$ goes into the nebula while the rest $(1 - \epsilon_{\rm out})$ is channeled into thermal X-ray emission. For maximum energy content in the nebula, we assume $\epsilon_{\rm out} = 1$.

Post energy injection, adiabatic losses play a prominent role in deciding the energy content at any given age. The expansion of the nebula and corresponding adiabatic cooling are detailed in the next subsection.

\subsection{Adiabatic losses}\label{adiabatic_loss}

The objective of this section is to determine the time evolution of the energy density in the magnetar wind nebula (MWN). To achieve this, we require the radius of the system and the energy content of the nebula at any given time. We present here a dynamical framework to describe both quantities.

To estimate the energy content of the nebula, we adopt the formalism of G17. At any given time, the energy content reflects the cumulative energy injected up to $t$, minus the adiabatic losses sustained by the injected energy. In the absence of adiabatic losses, the energy content is simply the time-integrated injection rate. However, when adiabatic losses are significant, the energy content is reduced compared to this cumulative value. The energy content at a given age, $t$, is expressed as:
\begin{equation}
    E_\mathrm{n}(t ) = \int_0^{t} \dot{E}_\mathrm{inj}(t') f_\mathrm{ad}(t') \, dt', \label{E_neb}
\end{equation}
where $f_\mathrm{ad}(t') = R(t') / R(t)$ is a dimensionless ratio that accounts for the adiabatic dilution of energy injected at earlier times as the MWN expands.  If most of the energy is injected early, it experiences greater adiabatic losses compared to scenarios where energy injection occurs predominantly at later times. A detailed summary of $f_\mathrm{ad}$ can be found in Table 3 of G17. 

\begin{figure}
    \centering
    \includegraphics[scale=0.55]{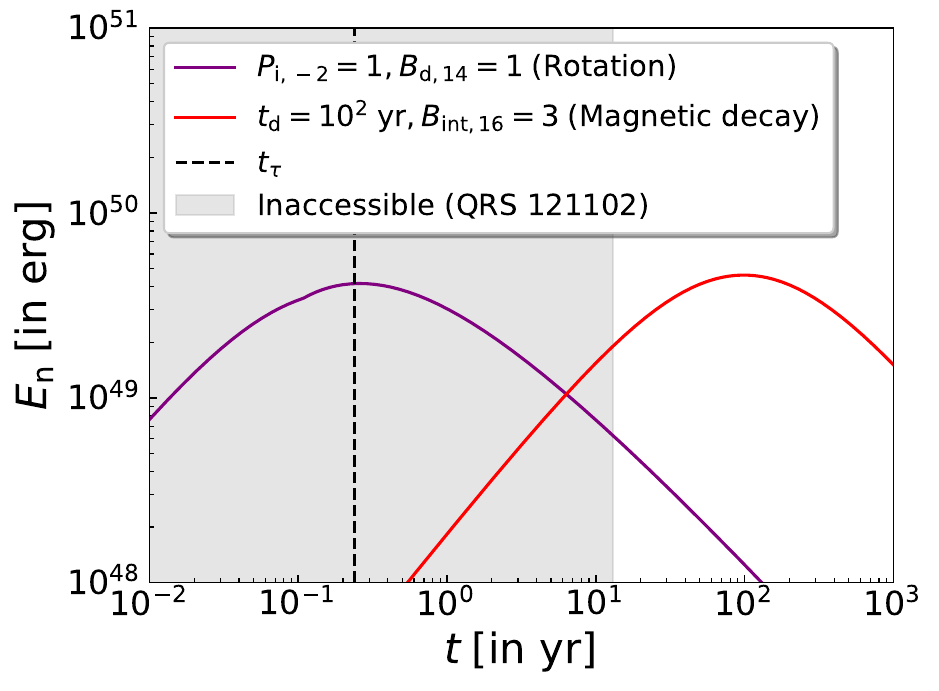}
    \caption{Illustrative energy content in the nebula powered by a magnetar as a function of time, taking into account adiabatic losses (see also Fig.~\ref{fig:switch_time_illustration} in appendix \ref{app:rot_mag_switch} for behavior of the dynamical energy injection).  The solid red line corresponds to the rotation powered MWN with  $P = 10\;$ms and $B_{\rm d} = 10^{14}\;$G.  The solid red line corresponds to the internal magnetic decay powered MWN with internal $B_{\rm int} = 3 \times  10^{16}\;$G with decay timescale $t_{\rm d} = 100\;$yr. The vertical dashed black  corresponds to the time $t_{\tau}$ (see equation \ref{Thomson_thin}) when the SNR ($M_\mathrm{SNR} = 3 M_{\odot}, E_\mathrm{SNR} = 10^{51} \; \text{erg}, f_\mathrm{ion} = 0.1$) becomes optically thin. The shaded gray region to the left of 13 years indicate that this region is not accessible to PRS associated with FRB 20121102A (the absolute minimum age for this source). }
    \label{fig:neb_en_cont}
\end{figure}

Figure \ref{fig:neb_en_cont} illustrates how the energy content in the nebula varies as a function of time. For any injection scenario, the energy content in the nebula increases, reaches a maximum near the characteristic timescale of the injection mechanism, and decreases beyond it (see appendix \ref{app:rot_mag_switch} for a detailed discussion). In particular, it can be seen that for the magnetic decay powered scenarios with an internal magnetic decay timescale $t_{\rm d} \geq 10\;$yr, the maximum available energy in the nebula is $E_{\rm n,max} \approx 4 \times 10^{49}$ erg , which is just twice the energy injected in the last dynamical timescale , $t_{\rm d}L_{\rm B}(t_{\rm d})$, (see Eq.~(\ref{max_enj}) in appendix \ref{app:rot_mag_switch}). The non-thermal electrons carry a fraction $\epsilon_{\rm e}$ of the nebular energy $E_{\rm n}$. In our analysis, to produce the most luminous PRS we require a highly efficient process with $\epsilon_{\rm e} = 1$.

Next, we can estimate the estimate the time at which the SNR becomes optically thin. The number density of free-electrons in the ejecta can be estimated from $n_{\rm e}  \approx  f_{\rm ion} M_{\rm SNR}/ 4 \pi R^3 m_\mathrm{p}$ where $f_{\rm ion}$ is the ionization fraction in the ejecta. The Thomson optical depth in the ejecta is given as $\tau_{\rm T} \approx n_{\rm e} \sigma_{\rm T} R$. Since the radius in the coasting phase scales linearly with time, we can estimate a critical time $t_{\rm 
\tau}$ where the remnant becomes optically-thin to Thomson scattering. It can be estimated as,
\begin{equation}
    t_{\tau}(\tau_{\rm T} = 1) \approx 0.24 \; f_{\rm -1} \;  M_{3} \; E_{\rm SNR,51}^{-\frac{1}{2}} \;  \text{yr}, \label{Thomson_thin} 
\end{equation}
where $f_{-1} = f_{\rm ion}/0.1$.  Appendix \ref{ap:free_free} discusses the constraints from free-free absorption. Adiabatic losses are discussed further in  \S\ref{MWN_param}.
 
\subsection{Constraints from PRS size and spectra}\label{spec_radio}

Next we describe the constraints on the age of a PRS from its observed size and spectrum.  In appendix \ref{app:SNR_flux} we demonstrate the the radio emission from the shock that the SNR drives into the ISM is expected to be negligible compared to that from the nebula, and cannot account for the PRS emission. This motivates us to consider MWN as the only viable source of synchrotron flux from the PRS. 

The non-thermal distribution of electrons in the MWN can be represented as
\begin{equation}
    \frac{{\rm d}N_e}{{\rm d}\gamma_{\rm e}}=(p-1) \frac{N_{\rm rel}}{\gamma_m} \left(\frac{\gamma_{\rm e}}{\gamma_m}\right)^{-p}  \hspace{0.2cm} \text{for $ \gamma_\mathrm{m} \leq \gamma_\mathrm{e} \leq \gamma_\mathrm{M} \;\& \; p \; >2$} , \label{non-thermal}
\end{equation}
where $N_{\rm rel}$ is the total number of non-thermal electrons in the nebula, $p$ is the power-law index, $\gamma_\mathrm{e}$ is the Lorentz factor of the non-thermal electrons and $\gamma_\mathrm{m}$ is the minimal Lorentz factor of the non-thermal electrons. For $p>2$, the average particle anergy in the non-thermal distribution is $\langle\gamma_e\rangle=\frac{p-1}{p-2}\gamma_\mathrm{m}\sim \gamma_\mathrm{m}$. For simplicity, we adopt a power-law index $p = 2.5$ for the nonthermal particle distribution.  The synchrotron frequency associated with electrons of Lorentz factor $\gamma_\mathrm{e}$ is defined as $\nu_\mathrm{syn} \equiv  \nu_\mathrm{B}\gamma^2_\mathrm{e}$ where the cyclotron frequency is defined $\nu_\mathrm{B} \equiv e B/2 \pi m_\mathrm{e} c$. Here $-e$  and $m_e$ are the charge and mass of the electron, and $c$ is the speed of light.

\begin{table*}
    \centering
    \caption{Derived parameters for the MWN. $B$ is the average magnetic field, $N_\mathrm{rel}$ is the total relativistic non-thermal electrons, and $E_\mathrm{rel}$ is the energy carried by $N_\mathrm{rel}$. We assume $p = 2.5$ and use $Q_\mathrm{x} = 10^\mathrm{x}$ for any quantity $x$ (in c.g.s units). Notation: $\gamma_\mathrm{m,2} = \gamma_\mathrm{m}/10^2$, $\nu_\mathrm{GHz} = \nu/(1\;\mathrm{GHz})$, $L_{\nu,28} = L_{\nu}/(10^{28}\;\mathrm{erg\;s^{-1}\;Hz^{-1}})$, and $R_{\rm 17} = R_{\rm n}/(10^{17}\;\mathrm{cm})$.}
    \begin{tabular}{|c|c|c|c|}
        \hline 
        Scenario & $B$ [mG] $\sim$ & $N_\mathrm{rel} \sim$ & $E_\mathrm{e}$ [erg] $\sim$ \\ 
        \hline 
        $\nu > \nu_\mathrm{m}$ & 
        $20 L_{\nu,28}^{\frac{4}{15}} \gamma_{\rm m,2}^{-\frac{2}{15}} \nu_{\rm GHz}^{\frac{1}{5}} \sigma^{\frac{4}{15}} R_{\rm 17}^{-\frac{4}{5}}$ & 
        $2 \times 10^{51} L_{\nu,28}^{\frac{8}{15}} \gamma_{\rm m,2}^{-\frac{19}{15}} \nu_{\rm GHz}^{\frac{2}{5}} \sigma^{-\frac{7}{15}} R_{\rm 17}^{\frac{7}{5}}$ &
        $4 \times 10^{47} L_{\nu,28}^{\frac{8}{15}} \gamma_{\rm m,2}^{-\frac{4}{15}} \nu_{\rm GHz}^{\frac{2}{5}} \sigma^{-\frac{7}{15}} R_{\rm 17}^{\frac{7}{5}}$ \\ 
        \hline
        $\nu < \nu_\mathrm{m}$ & 
        $10 L_{\nu,28}^{\frac{3}{8}} \gamma_{\rm m,2}^{\frac{5}{8}} \nu_{\rm GHz}^{-\frac{1}{8}} \sigma^{\frac{3}{8}} R_{\rm 17}^{-\frac{9}{8}}$ & 
        $10^{51} L_{\nu,28}^{\frac{3}{4}} \gamma_{\rm m,2}^{\frac{1}{4}} \nu_{\rm GHz}^{-\frac{1}{4}} \sigma^{-\frac{1}{4}} R_{\rm 17}^{\frac{3}{4}}$ & 
        $2 \times 10^{47} L_{\nu,28}^{\frac{3}{4}} \gamma_{\rm m,2}^{\frac{5}{4}} \nu_{\rm GHz}^{-\frac{1}{4}} \sigma^{-\frac{1}{4}} R_{\rm 17}^{\frac{3}{4}}$ \\
        \hline
    \end{tabular}\label{tab:der_spectra}
\end{table*}

Since MWN emits synchrotron radiation we apply the equipartition theorem \citep{1977MNRAS.180..539S,2013ApJ...772...78B}  to obtain an estimate of the equipartition radius and energy for the non-thermal electrons, which can serve as a proxy for the radius of the nebula for PRS systems. For a synchrotron self-absorption frequency that satisfies $\nu_{\rm sa} < \nu_{\rm m}=\nu_\mathrm{syn}(\gamma_\mathrm{m})$ (as observed), 
\begin{equation}
\begin{split}
     R_{\rm eq} \approx 2.4 \times 10^{17}  \;\nu_{\rm sa,8.7}^{-\frac{35}{51}} \; \nu_{\rm 9.3}^{-\frac{16}{51}}\; L_{29.3}^{-\frac{8}{17}} \; \text{cm}\; , \\
     E_{\rm eq} \approx 1.2 \times 10^{49} \; \nu_{\rm sa,8.7}^{-\frac{15}{17}} \; \nu_{\rm 9.3}^{-\frac{2}{17}}\; L_{29.3}^{\frac{20}{17}} \; \text{erg}\; , 
\end{split} \label{eq_radius_energy}
\end{equation}
where $\nu_{\rm sa,8.7} = \nu_{\rm sa}/(500\;$MHz), $\nu_{\rm 9.3} = \nu_{\rm m}/(2\;\text{GHz})$ and $L_{29.3} = L_{\nu_{\rm m}}/(2 \times 10^{29} \; {\rm erg\;s^{-1}\;Hz^{-1}})$,
which shows that for the PRSs associated with FRB 20121102A (and FRB 20190502B) and FRB 20201124A the equipartition radii are $\approx$ 0.1 pc and 1 pc respectively. The equipartition radius estimates for the PRSs associated with FRB 20190502B and FRB 20201124A are more constraining than upper limits based on radio imaging (see Table \ref{tab:properties}). Using the formalism from \citealt{2013ApJ...772...78B}, we can express energy in the nebula as $E_\mathrm{neb} = E_\mathrm{eq} \left[ \epsilon_\mathrm{B} \left( \frac{R}{R_\mathrm{eq}} \right)^{-6} + \epsilon_\mathrm{e} \left( \frac{R}{R_\mathrm{eq}} \right)^{11}  \right]$. For a maximum energy content of the nebula, the largest nebular radius consistent with all the observables satisfies the following relationship,
\begin{equation}
\begin{split}
     \frac{R_{\rm n,max}}{R_{\rm eq}} &\ \approx \left[ \frac{1}{\epsilon_{e}} \left( \frac{E_{\rm neb,max}}{E_{\rm eq}} \right) \right]^{\frac{1}{11}} \\
     &\ \approx 1.2 \; \epsilon_{\rm e}^{-\frac{1}{11}} \; \hspace{-0.2 cm} \hspace{0.2cm} \text{for $(E_{\rm neb,max},E_{\rm eq}) = (4,1.2) \times 10^{49}$ erg} , 
\end{split}
\end{equation}
where we have used the maximum nebular energy from internal magnetic field decay (see figure \ref{fig:neb_en_cont}).

\begin{figure}
\centering
\includegraphics[scale=0.55]{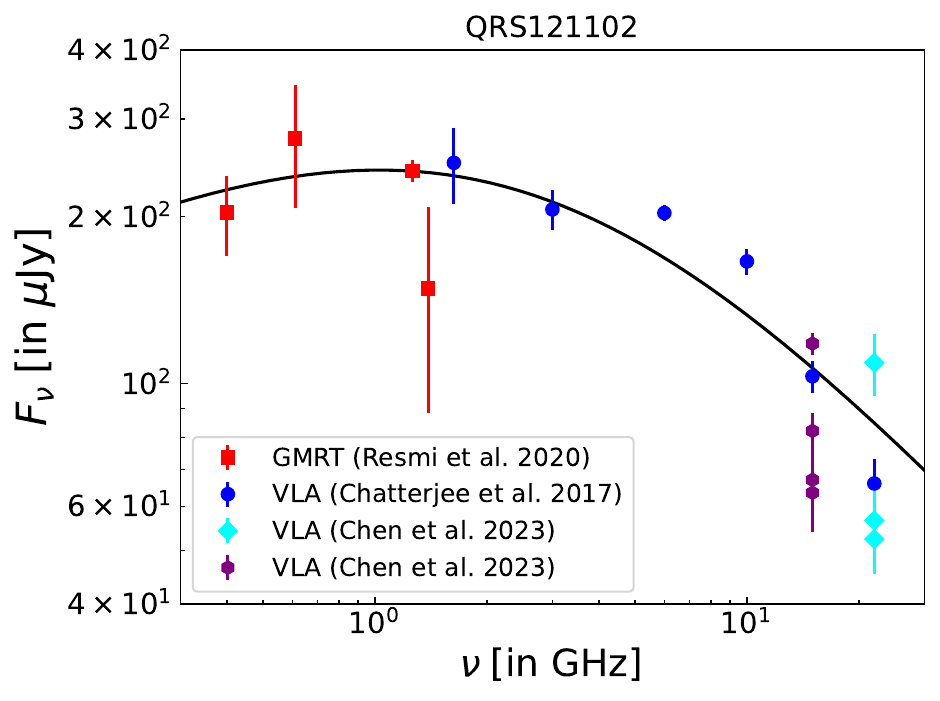}
\caption{The spectrum of the PRS associated with FRB121102 (a.k.a. QRS121102 in some studies). The red squares represent observations from GMRT (\citealt{Resmi21}), the blue circles represent the observations from VLA \citep{Chatterjee2017} and the purple pentagons and the cyan diamonds represent later observations demonstrating radio scintillations using VLA Ku-band (12\,--\,18\;GHz) and K-band (18\,--\,26\;GHz) (\citealt{2023ApJ...958..185C}). The black solid line represents a best fit model $(\nu_{\rm b} = 2.5\;  \text{GHz}, p =2.5)$ corresponding to the synchrotron slow cooling spectrum 1 of \cite{2002ApJ...568..820G}. The spectrum remains optically-thin even till 500 MHz while radio scintillation effects in later studies suggest the cooling break $\nu_{\rm c} > $ 22 GHz (see text for detailed discussion). } \label{fig:spec_PRS121102}
\end{figure}

Figure \ref{fig:spec_PRS121102} shows the PRS spectrum associated with FRB 20121102A. The shape of the spectrum shows that the self-absorption frequency $\nu_\mathrm{sa}$ lies below the peak frequency $\nu_\mathrm{m}$, while the cooling break frequency $\nu_\mathrm{c}$ is higher than $\nu_\mathrm{m}$ (corresponding to spectrum 1 in \citealt{2002ApJ...568..820G}). In general, $\nu_\mathrm{c}$ reflects the age of the system, while $\nu_\mathrm{sa}$ is indicative of its size. The size of the system $R$ is related to its age through equation \ref{neb_rad}.  The separation between $\nu_\mathrm{c}$ and $\nu_\mathrm{sa}$ is governed by the expansion history of the nebula. For a given nebular size, a younger system (with a smaller $t$) will exhibit a wider frequency separation, while an older system will show a narrower separation. For an observed spectrum $(\nu_{\rm sa}, \nu_{\rm m}, L_{\nu_{\rm m}}, \nu_{\rm c} )$ and  a fiducial  energy content of the nebula (assuming values for the PRS associated with FRB 20121102A) the radius $R_{\rm n}$ of the nebula can be estimated as (see appendix \ref{ap:neb_en})
\begin{equation}
     R_{\rm n} \approx 2 \times 10^{17} \; \nu_{\rm sa,8.7}^{-\frac{5}{7} } \; \nu_{\rm m,9.3}^{-\frac{5}{14}} \; L_{29.3}^{\frac{3}{7}} \; \nu_{\rm c,11}^{-\frac{13}{42}}\;  E_{50.3}^{\frac{1}{14}} \; \; M_{10}^{-\frac{1}{14}}\text{cm}\; ,\label{en_neb}
\end{equation}
 where $E_{\rm 50.3} = E_{\rm SNR} /(2\times10^{50}\;$erg) and $M_{10} = M/(10\;M_{\odot})$.
The age of the PRS can be estimated as (see appendix \ref{ap:neb_dyn} for a closure relation between the observables of the PRS)
\begin{equation}
t \approx 37 \; \nu_{\rm sa,8.7}^{-\frac{5}{7}} \; \nu_{\rm 9.3}^{- \frac{5}{14}} \; \nu_{\rm c,11}^{-\frac{13}{42}}\; L_{29.3}^{\frac{3}{7}} \; E_{50.3}^{-\frac{3}{7}} \; M_{10}^{\frac{3}{7}} \; \text{yr}\; ,\label{age_PRS}  
\end{equation}
 
Equation (\ref{age_PRS}) shows that characterizing the spectrum of the PRS can provide robust constraints on the age of the system. The PRS spectrum also allows the estimation of the energy of the non-thermal electrons $E_{\rm e}$, average magnetic field $B$, the total number of non-thermal electrons $N_{\rm rel}$ and minimal LF $\gamma_{\rm m}$ and cooling LF $\gamma_{\mathrm c}$ of non-thermal electrons as summarized below
 \begin{eqnarray}
 &\ E_{\rm e}\! \approx\! 1.4 \!\times\! 10^{49} \;     \nu_{\rm sa,8.7}^{\frac{5}{7}} \;  \nu_{\rm m, 9.3}^{\frac{1}{7}} \; \nu_{\rm c, 11}^{\frac{3}{7}}\;  L_{29.3}^{\frac{10}{7}}\; E_{50.3}^{-\frac{3}{7}} M_{10}^{\frac{3}{7}}\; \text{erg}, \\ 
     &\   B \approx 23 \;     \nu_{\rm sa,8.7}^{\frac{10}{21}} \;       \nu_{\rm m, 9.3}^{\frac{5}{21}} \; \nu_{\rm c, 11}^{-\frac{4}{21}}\; L_{29.3}^{-\frac{2}{7}}\; E_{50.3}^{\frac{2}{7}} M_{10}^{-\frac{2}{7}}\; \text{mG} ,                         \\
       &\ N_{\rm rel} \approx 2 \times 10^{52} \;  \nu_{\rm sa,8.7}^{- \frac{10}{21}}\;  \nu_{\rm 9.3}^{-\frac{5}{21}} \; L_{29.3}^{\frac{9}{7}} \; E_{50.3}^{-\frac{2}{7}} M_{10}^{\frac{2}{7}} , \\
    &\ \gamma_{\rm m} \approx 175 \;   \nu_{\rm sa,8.7}^{-\frac{5}{21}} \;      \; \nu_{\rm m, 9.3}^{\frac{8}{21}} \; \nu_{\rm c, 11}^{\frac{2}{21}}\; L_{29.3}^{\frac{1}{7}}\; E_{50.3}^{-\frac{1}{7}} M_{10}^{\frac{1}{7}}\; ,\\
    &\ \gamma_{\rm c} \approx 1223 \;  \nu_{\rm sa,8.7}^{-\frac{5}{21} }\;\nu_{\rm m,9.3}^{-\frac{5}{42}} \; \nu_{\rm c, 11}^{\frac{29}{42}}\;L_{ 29.3}^{\frac{1}{7}} \; E_{50.3}^{- \frac{1}{7}}\;  M_{10}^{ \frac{1}{7}}
 \end{eqnarray}
 
For the sub-energetic supernova parameters used above the SNR velocity $v_{\rm SNR} \approx 1.4 \times 10^{3} \;  E_{50.3}^{ \frac{1}{2}}\;  M_{10}^{ -\frac{1}{2}}$ km s$^{-1}$ is larger than the typical estimated transverse kick velocity of magnetars (see table 1 of G17). This is consistent with our assumed set-up  where the neutron star is well-bounded within the SNR as shown in Figure \ref{MWN-GF_setup}.  The requirement of a massive ejecta requires that the ionization fraction $f_{\rm ion}$ in the SNR ejecta be small so as not to exceed the DM contribution from the source. This can be expressed as   
\begin{equation}
    {\rm DM} \approx 343 \; \text{pc cm$^{-3}$} f_{\rm ion,-1.5} \; M_{10} \; R_{17.3}^{-2} \; ,
\end{equation}
where $f_{\rm ion,-1.5} =f_{\rm ion}/ 0.03 $. In our model, we treat the ionized fraction of the ejecta, $f_{\mathrm{ion}}$, as a free parameter. In the earlier work by \citet{Piro2016}, a constant ionization fraction was assumed, which led to a substantial variation in the dispersion measure (DM) over time. However, a more refined model by \citet{Piro18} allowed $f_{\mathrm{ion}}$ to evolve as a function of time/radius. This time-dependent treatment significantly reduced the expected DM variation, yielding a value of approximately $1~\mathrm{pc~cm^{-3}~yr^{-1}}$ at an age of $t = 40$~yr for an ejecta mass $M_1 = 10$ (see their Equation~17). This estimate is consistent with observational constraints on long-term evolution of DM in FRB 201211202A \citep{Spitler14,Jahns23}. Importantly, the estimate provided by \citet{Piro18} accounts only for ionization due to shock heating. Photo-ionization may also contribute significantly. For instance, photo-ionization modeling by \cite{Metzgar17} (see their Appendix) suggests that $f_{\mathrm{ion}}$ could range from 0 to 0.25. Given these considerations, we choose to keep $f_{\mathrm{ion}}$ as a free parameter in our analysis. We further constrain $f_{\mathrm{ion}} \leq 0.03$, based on the requirement that the DM contribution from the SNR does not exceed the total observed DM.

If the full shape of the spectra is not known i.e., only the spectral luminosity $L_{\nu}$ at a given frequency $\nu$ is measured, rough estimates of few fiducial properties can still be made as shown in Table \ref{tab:der_spectra}.

In \S \ref{MWN_param} we show what kind of magnetar parameter space can account for the necessary MWN energetics to power these PRS sources at their current age.

\section{Constraining the rotation/magnetic powered MWN parameter space}\label{MWN_param}
We explore next the rotational and magnetic power parameter space for magnetar wind nebulae (MWN) capable of producing persistent radio sources (PRSs). The key question is identifying the conditions under which the system can sustain its current radio luminosity and compact size. The predicted features in the PRS spectrum can be tested by observations to distinguish which among the two channels is powering these sources or ruling out the magnetar interpretation altogether. To ensure that the internal magnetic field satisfies \( B_\mathrm{int} \geq 10^{15} \, \text{G} \) -- a necessary condition for FRB production (see \S \ref{req_FRB}) -- we focus on dipolar magnetic field strengths in the range \( B_\mathrm{d} = 10^{14} - 10^{15} \, \text{G} \) (\S \ref{eng_inj}).  

The evolution of the nebula is governed by six critical timescales: the system age $t$, which determines its size (Eq.\ref{neb_rad}); the optical depth timescale $t_{\tau}$, when the expanding ejecta becomes optically thin to Thomson scattering (Eq.\ref{Thomson_thin}); the spin-down timescale $t_\mathrm{sd}$, over which the magnetar loses most of its rotational energy (Eq.\ref{rot_parm}); the switching timescale $t_{\rm switch}$, marking the transition from rotational energy loss to internal magnetic field decay (Eq.\ref{time_switch}); the magnetic field decay timescale $t_\mathrm{d}$, representing significant depletion of internal magnetic energy (Eq.\ref{mag_param}); and the Sedov-Taylor time  $t_\mathrm{dec} $, indicating the onset of SNR ejecta deceleration (Eq.~\ref{time_ST}). The relevance of $t_\mathrm{dec}$ depends on whether it is shorter than the other timescales. At age $t$, the dominant energy source of the nebula is determined by $ t_{\rm switch}$: if $t_\mathrm{sd} < t < t_\mathrm{switch}$, rotational energy losses dominate, while if $t_\mathrm{switch} < t < t_\mathrm{d}$, internal magnetic field decay powers the nebula (see appendix \ref{app:rot_mag_switch} for trends).

Any scenario that powers the MWN+SNR system must satisfy the following stringent conditions:
\begin{enumerate}
\item If the radio spectrum shows no sign of steepness till the lowest frequency at which the PRS has been observed (see Fig.~\ref{fig:spec_PRS121102}) it implies that the self-absorption frequency must be lie at even lower frequency. 
\item The cooling break must be higher if no cooling break has been observed even at the highest observed frequency. 
\item The source cannot be younger than the duration over which it has been observed as a PRS.
\item The DM contribution from the source cannot exceed the total measured value.
\end{enumerate}

\begin{figure*}[ht]
\centering
\begin{minipage}[b]{0.4\textwidth}
\begin{minipage}[b]{\textwidth}
\includegraphics[scale = 0.46]{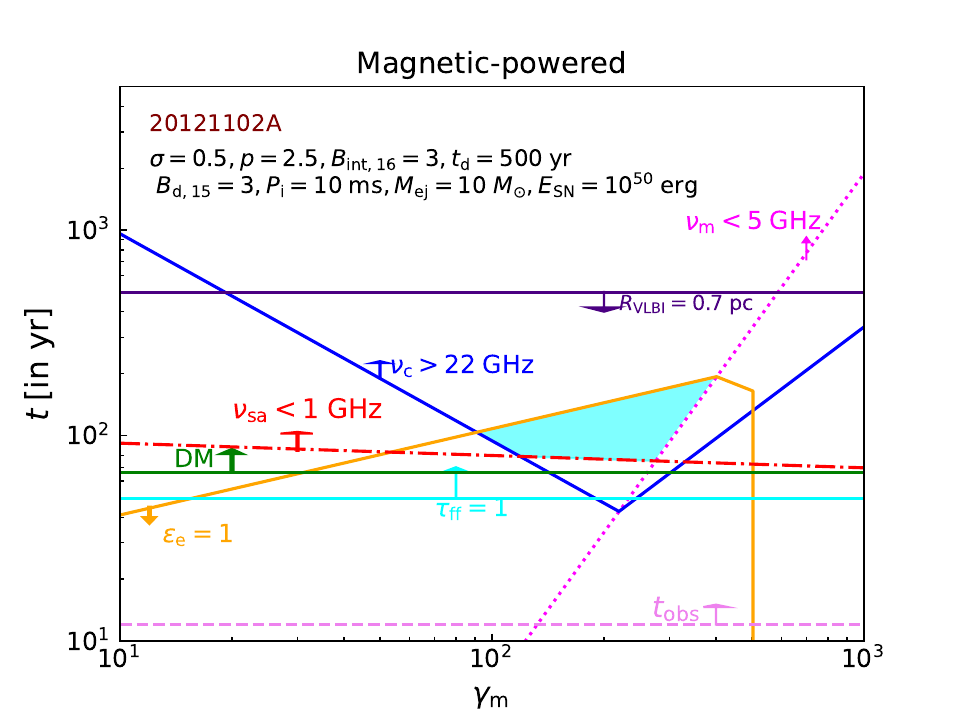}
\end{minipage}
\vspace{\baselineskip}
\begin{minipage}[b]{0.4\textwidth}
\includegraphics[scale=0.46]{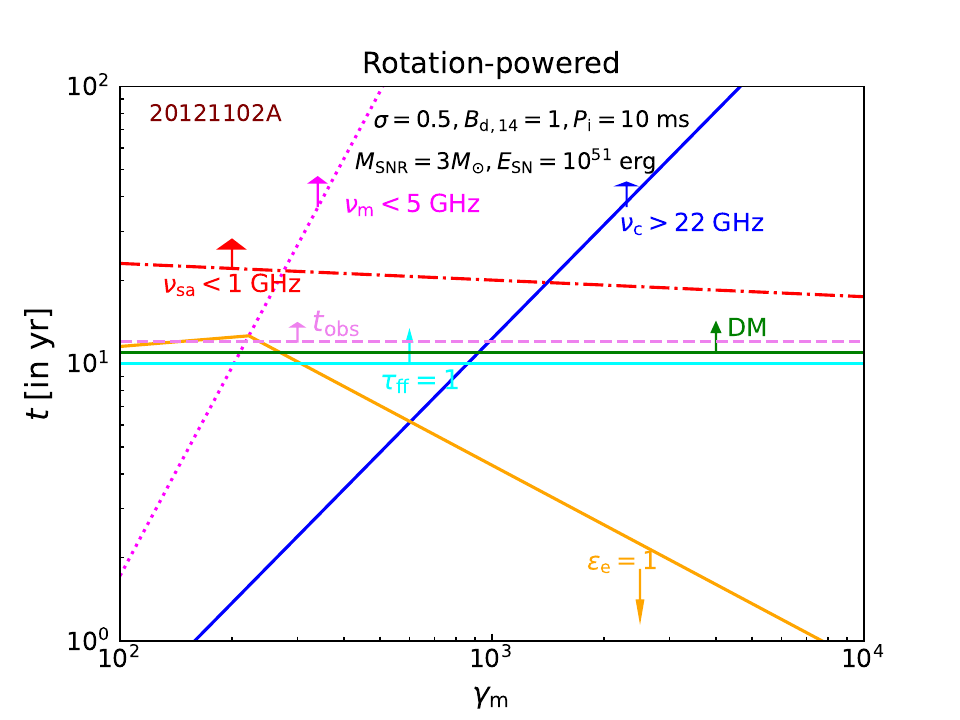}
\end{minipage}
\end{minipage}
\hfill
\begin{minipage}[b]{0.5\textwidth}
\hspace{-1cm}
\includegraphics[scale=0.50]{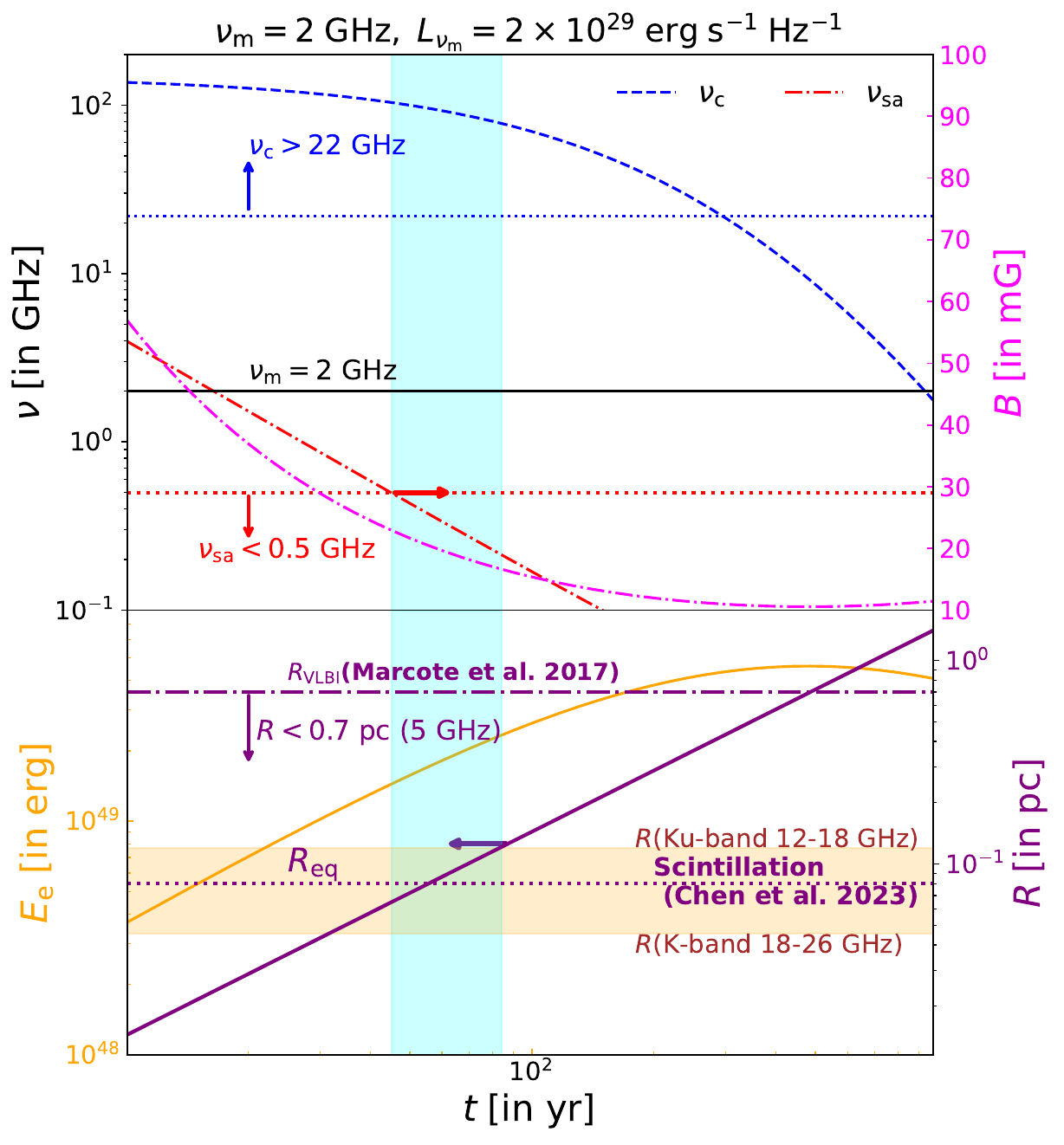}
\end{minipage}
\caption{Rotation/Magnetic-decay powered parameter space for PRS associated with FRB 20121102A. The parameter values used are: $\epsilon_\mathrm{e} =1, \epsilon_\mathrm{out} = 1,\sigma = 0.5, p = 2.5 , , P_{\rm i} = 10 \; \text{ms}, B_{\rm d,14} = 1, B_{\rm int,16} = 3$.    \textbf{Left:} \textit{Top Left panel:} shows the allowed region (in shaded cyan) of magnetar-powered MWN in the $t - \gamma_{\rm m}$ plane with magnetic decay timescale $t_{\rm d} = 500$ years for a sub-energetic supernovae embedded in massive ejecta $M_\mathrm{ej} = 10\; M_{\odot}$. The constraints used are: $t_{\rm obs} > 13\;$yr (horizontal pink dashed line), $\nu_{\rm c}>22\;$GHz (blue), $\nu_{\rm sa}<1\;$GHz (dot-dashed red line), the age must be large enough not to exceed the on-source DM (solid green line), the source size should be smaller than the upper limit suggested by imaging studies (solid purple line), the energy content of the non-thermal electrons cannot exceed the total energy content in the nebula (solid yellow). \textit{Left bottom panel:} shows no parameter space exists for the rotation-powered parameter space. All the lines have the same meaning as in the previous panels. \textbf{Right:}  shows the allowed parameter space in the magnetic decay-powered space in the blue strip ( for the same magnetar progenitor and SN property as shown in the top left panel) for a fixed observable $\nu_{\rm m} = 2$ GHz with $L_{\nu_{\rm m}} = 2 \times 10^{29}$ erg s$^{-1}$ Hz$^{-1}$. The radius constraint from scintillation studies is from \cite{2023ApJ...958..185C} and its upper limit from imaging studies is taken from \cite{Marcote2017}. The equipartition radius (see  Eq.~\ref{eq_radius_energy}) is shown as a purple horizontal dotted line in the lower subpanel. In the upper subpanel, the red arrow pointing rightwards shows the left boundary of the blue strip corresponding to a lower limit on the nebular age of 45 years is constrained by the self-absorption frequency. In the lower subpanel, the purple arrow pointing leftwards shows the right boundary of the blue strip corresponding to an upper limit on the nebular age of 88 years, which is constrained by the upper limit on the nebular radius from radio scintillation studies ( \citealt{2023ApJ...958..185C} where they assumed a scattering screen at a distance of 1 kpc from the source).\label{fig:Mag_121102} }
\end{figure*}

\begin{figure}
\centering
\begin{tabular}{c}
\includegraphics[scale=0.53]{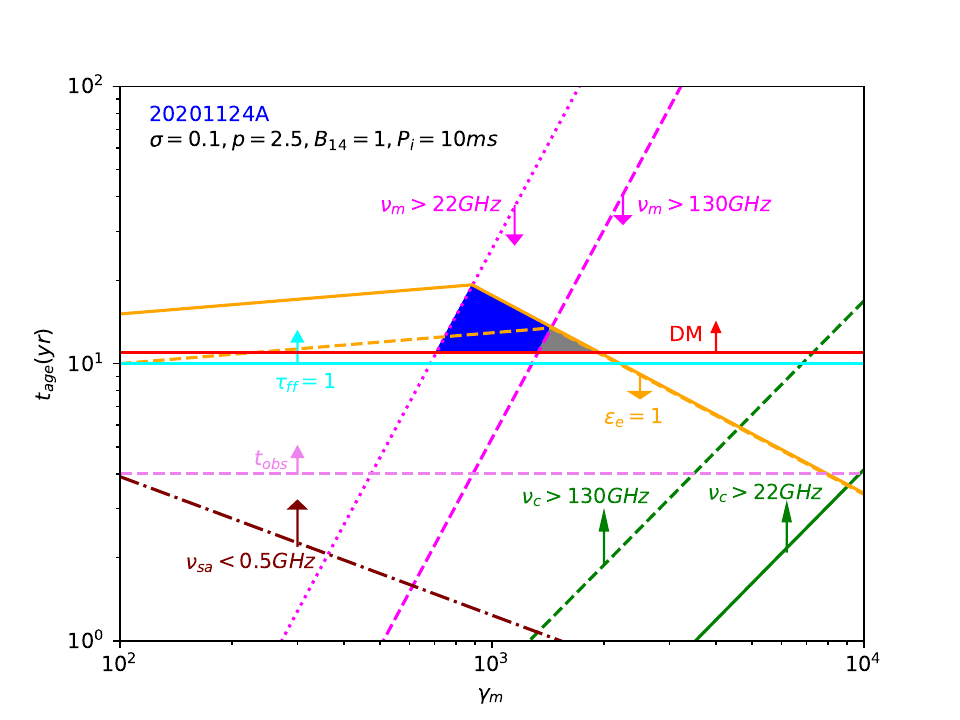}  \\
\hspace{-1cm}\includegraphics[scale=0.51]{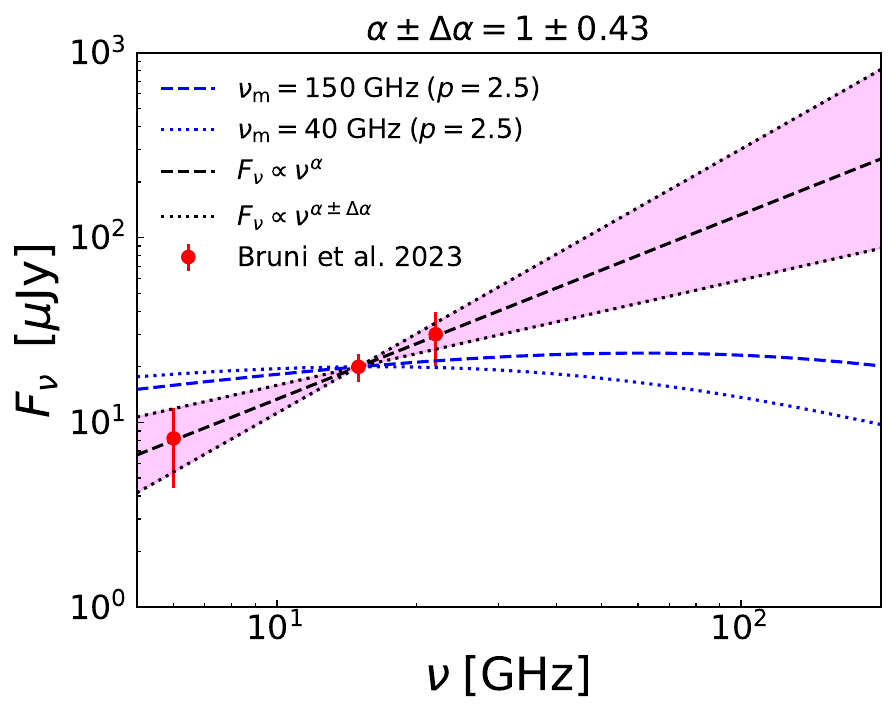}
\end{tabular}
\caption{Rotation-powered powered parameter space for candidate PRS associated with FRB 20201124A. The parameter values used are: $\sigma = 0.1, p = 2.5, P_{\rm i} = 10 \; \text{ms}, B_{\rm d,14} = 1, M_{\rm SNR} = 3 M_{\odot}, E_{\rm SNR} = 10^{51} \; \text{erg}$.\textbf{Top:} The shaded region shows the allowed parameter space in the $t - \gamma_{\rm m}$ plane. The different lines have the same meaning as in Figure \ref{fig:Mag_121102}. \textbf{Bottom:} shows the inverted spectrum of the source and the possible turnover at higher frequencies shown in dashed blue thick line ($\nu_{\rm m} = 150$ GHz) and thin line ($\nu_{\rm m} = 40$ GHz). The red data points at 6, 10 and 22 GHz along with the error bar on the spectral index are taken from \cite{2023arXiv231215296B}.\label{fig: Rot_201124}}
\end{figure}

For both channels, we assume an initial spin period $P_{\rm i} \gtrsim10\;$ms (for which $E_{\rm rot} \lesssim 2 \times 10^{50}$ erg). Under this assumption, the nebula size at a given time, $R_\mathrm{n}(t)$, is determined by the ratio of the supernova explosion $E_{\rm SN}$ and the ejected mass $M$.    

For rotational energy to effectively power the nebula, one can estimate the maximum dipolar magnetic field that can sustain the PRS luminosity $\nu L_{\nu}$ at age $t$ which can be estimated (assuming magnetic dipole breaking) as  (see appendix \ref{max_rot_Bd} for derivation) 
\begin{equation}
    B_{\rm d,max,rot} \approx   4 \times 10^{14}\;  \epsilon_{\rm e}^{\frac{1}{2}} (\nu L_{\nu})_{38}^{-1} \;  t_{10}^{-1} \;  \text{G}  \; \hspace{0.1cm} \text{for}\ m =2 \;,\label{max_Bdrot}
\end{equation}
where $(\nu L_{\nu})_{38} = \nu L_{\nu}/(10^{38}\, {\rm{}erg/s)}$ and $t_{10} = t /(10\;$yr).  

However, for the two confirmed PRSs the self-absorption frequency $\nu_{\rm sa}$ drops below 1 GHz only for $t \gtrsim 30\;$yr. Comparing Eqs.(\ref{Bd_frb}) and (\ref{max_Bdrot}), for $\nu L_{\nu} \gtrsim 4 \times 10^{38}\,{\rm erg\;s^{-1}}$ at $t \gtrsim 30\;$yr, it can be seen (even assuming $\epsilon_{\rm e} \rightarrow 1$) that the maximum dipolar magnetic field supporting PRS luminosity is lower than the minimum dipolar magnetic field required for FRB production. Thus, the two confirmed PRS–FRB sources cannot be powered by rotational energy loss for ages $t\geq 30$ years, necessitating internal magnetic field decay as the only viable alternative energy source. The rotational channel can sustain the PRS  luminosity and FRB production for only a very small window of around two decades.   However, for the third and much dimmer candidate PRS ($\nu L_{\nu} \sim 10^{36}$ erg s$^{-1}$), the rotational channel remains viable with $B_{\rm d} \sim 10^{14}$ G at $t_{10} = 1$.

For the magnetically powered case, there is a characteristic time $t_{\rm switch}$ beyond which the dominant source of energy injection into the nebula switches from rotational energy loss to internal magnetic field decay. This switching time $t_{\rm switch}$ can be expressed as (see appendix \ref{app:rot_mag_switch} for derivation), 
\begin{equation}
    t_{\rm switch} \approx 3.3 \; \epsilon_{\rm out}^{-\frac{1}{2}}\;t_\mathrm{d,3}^{\frac{1}{2}} \; f_\mathrm{dip,-2}^{-\frac{1}{2}} \;  B_{\rm int,16}^{-2 - \frac{\alpha_{\rm B}}{2}} \; \text{yr} \hspace{0.2cm} \text{for $\alpha_{\rm B} =1$}\;, \label{time_switch}
\end{equation}
where the  $f_{\rm dip,-2} = f_{\rm dip}/10^{-2}$ corresponds to our fiducial choice $f_{\rm dip}^{1/2}=\frac{B_{\rm d}}{B_{\rm int}} = \frac{1}{10}$. The energy content in the nebula and its radius at this time are 
given by (assuming $\alpha_{\rm B} = 1$),
\begin{equation}
\begin{split}
     &\ E_{\rm switch} = 2 \epsilon_{\rm out} \; E_{\rm B,0} \left(\frac{t_{\rm switch}}{t_{\rm d}} \right) \\
     &\ \approx 1.3 \times 10^{47} \; \; \epsilon_{\rm out}^{\frac{1}{2}}\;t_\mathrm{d,3}^{-\frac{1}{2}} \; f_\mathrm{dip,-2}^{-\frac{1}{2}} \;  B_{\rm int,16}^{1+ \frac{\alpha_{\rm B}}{2}} \; \text{erg} \hspace{0.5cm} , \\
     &\ R_{\rm switch} = v_{\rm SNR} t_{\rm switch} \\
     &\  \approx 5 \times 10^{16} \;  M_{\rm 3}^{-\frac{1}{2}} E_{\rm SNR,51}^{\frac{1}{2}}  \epsilon_{\rm out}^{-\frac{1}{2}}\;t_\mathrm{d,3}^{\frac{1}{2}} \; f_\mathrm{dip,-2}^{-\frac{1}{2}} \;  B_{\rm int,16}^{-2 - \frac{\alpha_{\rm B}}{2}} \;  \text{cm}. 
\end{split}
\end{equation}

Equation (\ref{time_switch}) indicates that a typical magnetar progenitor capable of powering both FRBs and a luminous MWN remains primarily in the rotation-powered phase for only about a decade. The duration of this transition phase is largely determined by the strength of the internal magnetic field, $B_{\rm int}$. We revisit the feasibility of detecting such a phase in \S \ref{obs_feas}. For $t>t_\mathrm{switch}$, the internal magnetic field decay channel dominates and the optimal scenario occurs when the internal magnetic field reaches its maximum allowable value $B_\mathrm{int,max} \sim 3 \times 10^{16}$ G with $\epsilon_\mathrm{out} \rightarrow 1$. Assuming the scaling in Eq.~(\ref{mag_param}) this condition likely corresponds to the shortest decay timescales $t_\mathrm{d} \sim10^{2.5}\;$yr and the largest possible dipolar magnetic field strength $B_\mathrm{d} \sim 10^{15}$ G, which consequently results in a very short spin-down timescale. As a result, the spin-down time is consistently shorter than the magnetic decay timescale. The initial spin period plays a negligible role, provided that the rotational energy does not exceed the supernova explosion energy. Even under these conditions, a more compact nebula can be achieved by increasing the ejecta mass or considering a weak/sub-energetic  supernova.

Figure \ref{fig:Mag_121102} illustrates that the rotation-powered parameter space is excluded for FRB 20121102A (and similarly for FRB 20190520B, not shown). This is because, beyond 30 years, rotational energy alone cannot sustain the high energy content required to power the PRS. The SN and initial magnetar birth parameters for this figure are chosen to give the longest possible age estimate $\sim $ 100 years for an internal magnetic field strength $B_{\rm int,16} = 3$ with a maximum decay timescale of $t_{\rm d,max} \sim 500$ years (see discussion in \S\ref{PRS_age}). The predicted cooling frequency in the PRS spectrum lies around $\nu_{\rm c} \sim 70-100$ GHz (which can be observed in the ALMA band; \citealt{ALMA2009}), while the self-absorption frequency is estimated to be $\nu_{\rm sa} \sim 200-500$ MHz (which can be observed in the LOFAR band (see discussion in \S \ref{predic_spec}); \citealt{2013A&A...556A...2V}). The magnetic field strength within the nebula is found to be in the range $10 \;  \text{mG} < B < 30 \;  \text{mG}$, consistent with the results of \cite{2022MNRAS.510.4654B} (see their Eq. 59). 

Figure \ref{fig: Rot_201124} shows that a rotation powered MWN is allowed for the candidate PRS associated with FRB 20201124A. However, this scenario only works if the system is very young ($\leq$ 20 years). This scenario can be ruled out if the inverted nature of the PRS spectrum persists beyond 150 GHz. This turnover can be observed in the ALMA band.

\section{Comparison with previous studies}\label{rob_conc}
\noindent
\textbf{Confirmed PRSs for FRBs 20121102A and 20190520B:}  

Our interpretation of the mechanism that powers PRS associated with FRB 20121102A aligns with the models proposed by \citet{Beloborodov2017} (hereafter B17) and \citet{MM2018} (hereafter MM18), but there are important differences that warrant further discussion. Both studies do not consider the impact of the SNR ejecta on the nebular properties. In their models, both the initial rotational period $P_{\rm i}$ and the total mechanical kinetic energy imparted to the SNR $(E_{\rm SN} + E_{\rm rot})$ cannot be constrained. In our model, we show that these magnetic powered systems must necessarily be produced in sub-energetic supernovae ($E_\mathrm{SN}\,\!\sim\,\!10^{50}\;$erg) with massive ejecta ($M \gtrsim 10 M_{\odot}$) and large initial rotation periods ($P_{\rm i}\geq 10\;$ms)  in order to satisfy the size constraints from observations. B17's estimate of the energy content of non-thermal electrons $E_{\rm e}$ is an order of magnitude smaller than ours, since they considered the cooling break to be at $\nu_{\rm c} \sim 10\;$GHz. This was before later studies (\citealt{2023ApJ...958..185C}) revealed significant scintillation effects, showing that $\nu_{\rm c}$ remains undetected even up to 22\;GHz. Consequently their estimate of the mean nebular magnetic field is $\sim 3$ times higher than in our analysis. 

MM18 suggested a wide parameter space associated with a millisecond pulsar for the age of these systems viz., $t  \sim 10-10^{3}\;$yr and radius $R \sim 0.1 - 1$ pc. In contrast, we find a very tight constraint on the age -- that these systems can at most be a century old. We also find a millisecond magnetar to be untenable as such a system will violate the observational constraint on radius. Further, MM18 assumed the magnetic decay timescale to be on the order of a year.  Since the magnetic decay time is assumed to scale as $B_{\rm int}^{-\alpha}$ \citep{Colpi2000,2012MNRAS.422.2878D}, such a short decay timescale implies a much stronger internal magnetic field strength (see Eq.~(\ref{mag_param})) far exceeding the maximum allowable value ($B_\mathrm{int} \gg B_\mathrm{int,max}$), which appears physically implausible.  
The requirement for a decay timescale of less than a year arises from the fact that FRB 20121102A is one of the most active repeating FRBs and MM18 makes use of the outer magnetospheric blast wave model \citep{2019MNRAS.485.4091M}. In the blastwave model, FRBs are produced by the interaction of giant flare ejecta with the magnetar wind. This model necessitates the ejection of multiple giant flares on short timescales to sustain a high repetition rate. Besides, MM18 assumed fixed supernova ejecta velocities and neglected the SNR to the DM, leading to an uncorrelated treatment of RM and DM -- contradicting observational evidence (see \S \ref{RM_DM_correlation}).

An alternative scenario, proposed by (\citealt{Kashiyama17,2024arXiv241219358B}, henceforth B24), suggests that the PRS of FRB 20121102A could be powered by rotational energy, assuming a dipolar magnetic field strength of $B_\mathrm{d} \leq 10^{13}$ G ( see appendix \ref{app:B24_Param} for the implications of the proposed parameter space by B24). In their model, PRS emission is driven by a neutron star with a millisecond spin period, yielding a substantial rotational energy reservoir of $E_\mathrm{rot} \approx 10^{52}\;$erg (see Eq.~(\ref{rot_parm})). The low magnetic field strength prolongs the spin-down timescale to $t_\mathrm{sd} \sim 1\;$yr  
(see Eq.~(\ref{rot_parm})), and from this point onward the released $E_\mathrm{rot}$ is converted by adiabatic cooling into the MWN+SNR expansion. After a decade the nebula already expands beyond 2 pc. Besides, this low $B_\mathrm{d}$ implies an internal magnetic field of $B_\mathrm{int} <  10^{13}\;$G, which is two orders of magnitude smaller than the minimum value required to produce FRBs (see \S\ref{req_FRB} for details). Moreover, low dipolar and internal magnetic fields presents additional challenges. In such environments, baryon-loaded outflows are less likely, complicating the explanation of the observed dispersion measure (DM-RM) correlation (see \S\ref{RM_DM_correlation} for discussion within the framework of our magnetically powered model). It is also more challenging to channel magnetic field decay energy into the MWN, which requires outflows associated with bursting activity. Thus, the rotation-powered model  fails to simultaneously account for both the FRB and the PRS associated with these sources.

Previous studies \citep{Murase16,Omand18,Murase21} have modeled the energy evolution of non-thermal electrons and nebular magnetic energy using different approaches. In their frameworks, while the energy of non-thermal electrons is subject to adiabatic losses, the magnetic energy of the nebula is assumed not to decay. This is implemented by setting the adiabatic loss coefficient for the magnetic field, $c_\mathrm{B}$, to zero. Consequently, the magnetic energy at any given time $t$ is taken to be a constant fraction, $\epsilon_\mathrm{B}$, of the total energy injected into the nebula up to that time. This modeling approach is motivated by spectral energy distribution (SED) studies \citep{Tanaka10,Tanaka11} of Galactic pulsar wind nebulae (PWNe), where adiabatic losses are typically neglected. As a result, the model inherently assumes a highly magnetized nebula, characterized by a large magnetization parameter, $\sigma$. However, this high-$\sigma$ scenario stands in tension with observational indications that the current nebular size is near the equipartition radius, where the magnetization is expected to be closer to 
$\sigma_\mathrm{eq} \sim 0.3 < 1$, 
suggesting a system closer to equipartition rather than one dominated by magnetic energy. An additional point for these previous studies is that, for the electron energy distributions, the cooling Lorentz factor is assumed to be very high. Consequently, these studies predict an absence of a cooling break in the 100 GHz range, which is exactly the range where we predict the cooling break should lie for these systems. In our analysis, we treat both the non-thermal electron energy and magnetic energy as undergoing adiabatic losses—namely, both energies are modeled as a fraction of the current nebular energy content, rather than the cumulative energy content for the magnetic energy. Since the non-thermal electrons dominate the energy content of the nebula, our model is inherently a low-magnetization model that satisfies equipartition estimates.

\smallskip\noindent
\textbf{Candidate PRS associated with FRB20201124A:}  The rotation-driven model is applicable here due to the lower radio luminosity, but the system must be very young (on the order of a few decades). For the magnetic decay model to be applicable for this source the internal magnetic decay timescale needs to be around few decades (i.e. three orders of magnitude smaller than what is inferred from galactic magnetars; see \citealt{2012MNRAS.422.2878D,Beniamini2019} ). The system's age can be determined by a turnover in the spectrum. Observing this source at sub-mm wavelengths is crucial: if no turnover is observed up to 180\;GHz, the rotation-powered MWN model can be ruled out.

\section{Discussion and conclusions}\label{conc}

The association of PRSs with repeater FRBs provides a new opportunity to test and constrain a magnetar wind nebula (MWN) origin for these sources. Producing compact, bright PRSs with ages between a few decades to a few hundred years requires a high energy density of non-thermal particles. However, adiabatic losses reduce the MWN's energy with time, while its expansion also further reduces its energy density. 
A high energy density at present requires a relatively late energy injection, where most of the energy is supplied close to the system’s current age. We have explored the magnetar parameters necessary for achieving this within rotational or magnetic energy-powered scenarios. Our comprehensive analysis shows that rotational energy alone cannot account for the observed luminosities and spectra of the two confirmed PRSs associated with FRBs 20121102A and 20190520B. Instead, we find that magnetar systems with moderate initial spin periods and very strong internal magnetic fields, with decay timescales $\sim10^{2.5}\;$yr, 
can successfully reproduce the observed radio properties. 

In the following subsections we discuss our findings and suggest observational strategies to corroborate our interpretation.

\subsection{Constraining the age and powering-mechanism}\label{PRS_age}

The age of the nebula is a critical factor in identifying the powering channel of the nebula. As discussed in \S\ref{MWN_param}, the rotation channel can sustain FRB production and observed PRS luminosity $\nu L_{\nu} \sim 10^{39}$ erg s$^{-1}$ only if the source is less than two decades old. Beyond this, only the magnetic-decay channel remains viable. However, the decay timescale for the internal magnetic field needs to be close to the current age to account for the observed radio luminosity. This in turn suggests that only for the oldest age of the system the slowest decay channel can be accommodated. 

Given the sub-parsec nature of the PRS, it is very likely that the SNR is in the coasting phase, suggesting that the radius of the nebula is equal to the radius of the remnant. The age of the nebula $t$ must be at least equal to the time for which the source has been observed $t \gtrsim t_\mathrm{obs} = 13$ years. For a given size of the nebula, the current age of the PRS source is inversely related to the expansion velocity of the nebula,
\begin{equation}
\begin{split}
     t = &\  \frac{R_\mathrm{SNR}}{v_\mathrm{SNR}} =  R_{\rm SNR} \sqrt{\frac{M_\mathrm{SNR}}{2 (E_\mathrm{rot} + E_\mathrm{SN})}} \\ 
     &\ \approx 
    \begin{cases}
        &\   2.5 \; R_{17.3} \; M_{3}^{\frac{1}{2}} \; P_{\rm i,-3}\; \text{yr} \hspace{0.9cm}  \text{for } E_\mathrm{rot} > E_\mathrm{SN},  \\
        &\ 63.4 \; R_{17.3} \;  M_{10}^{\frac{1}{2}} \; E_{50}^{-\frac{1}{2}} \; \; \text{yr}   \hspace{0.7cm}    \text{for } E_\mathrm{rot} < E_\mathrm{SN}. \\
    \end{cases} \label{age_PRS_range}
\end{split}
\end{equation}

Equation \ref{age_PRS_range} shows that for a millisecond magnetar candidate, the sub-parsec size is achieved at a very young age as the rotational energy loss boosts the expansion velocity of the nebula to a very high value while the initial supernova explosion energy remains sub-dominant. Since this age is smaller than \(t_\mathrm{obs}\), this makes it a very unlikely scenario. This also rules out the combination of an SLSN and a millisecond magnetar scenario as suggested by \citep{Metzgar17} (hereafter M17) where the initial explosion energy \(E_\mathrm{SN} \gtrsim 4 \times 10^{51}\) erg.  The unfeasibility of the latter scenario is also consistent with the absence of FRBs in deep radio surveys of SLSN at least a decade after the explosion \citep{Eftekhari21}

The highest expansion velocity (and the youngest inferred age $\sim$ 0.4 years for a sub-parsec size) is achieved for ultra-stripped supernovae (USSN) with mass \(M \sim 0.1 M_{\odot}\) with initial spin period \(P_\mathrm{i} = 1\) ms. This then makes the scenario preferred by B24 an unlikely scenario to form a PRS (see \S \ref{rob_conc}).

The second line of equation \ref{age_PRS_range} shows that only a subenergetic (weak) supernova with a massive ejecta and a slow-spinning central magnetar can provide the longest age of the system $\sim 100$ years. In this scenario, the rotation channel is quenched, and only the internal magnetic decay channel is viable. To support this scenario, we examined the maximum possible decay timescale $t_{\rm d,,max}$ (i.e. the slowest internal magnetic field decay) that can sustain the observed PRS luminosity. The conversion of the magnetic energy decay to the nebular energy itself needs to be very efficient. This can be understood as follows. The radius estimate from radio scintillation is close to the equipartition radius (see Fig.~\ref{fig:Mag_121102}), suggesting an energy content comparable to the equipartition value, $E_{\rm e} \geq 2 \times 10^{49}$ erg (see Eq.~(\ref{eq_radius_energy})). As the maximum energy in non-thermal electrons in the MWN is $5\times10^{48}\;\epsilon_{\rm e}\epsilon_{\rm out}B_{\rm int,16}^2$ erg (see appendix \ref{app:rot_mag_switch}), an internal field strength close to the maximum value of $B_{\rm int,16} \approx 3$ ) as well as $\epsilon_{\rm e}\epsilon_{\rm out}\sim1$ is required for powering the PRS.

In summary, the PRS must be as old as possible to accommodate gradual internal field decay. However, the PRS's sub-parsec size sets a strict limit on its expansion speed (which in the coasting phase matches that for the SNR), favoring a slow expansion speed and hence a weak explosion with a massive ejecta. The combination of a weaker explosion, a longer spin period, and a massive ejecta extends the system's age to hundreds of years, allows for the longest magnetic decay timescale. This in turn makes the sub-energetic SN the most compelling candidate for explaining the PRS. \textit{Any other combination for 
$(E_\mathrm{SN}, P_\mathrm{i},M_\mathrm{SNR})$ would only reduce the MWN's current age and require a very precise fine tuning to achieve a sub-parsec size.}

\subsection{Predictions for the PRS spectra}\label{predic_spec}

For the sub-energetic supernova scenario, the PRS can be at most a century old ($t \lesssim 100\;$ yr), and our analysis provides a lower limit on the mean MWN magnetic field, $B \gtrsim 10\;$mG (consistent with  \citealt{2022MNRAS.510.4654B}). The electron cooling Lorentz factor, $\gamma_{\rm c} = \frac{6 \pi m_{\rm e} c}{\sigma_{\rm T} B^2 t}$, corresponds to a cooling break frequency $\nu_{\rm c} \equiv \gamma_{\rm c}^2 \nu_{\rm B}$ ($\nu_{\rm B}$ being the cyclotron frequency) given by:
\begin{equation}
    \nu_{\rm c} \approx 170 \; B_{-2}^{-3} \; t_{100}^{-2}  \; \text{GHz}\;,\label{cool_break}
\end{equation}
where $B_{-2} = B/(10^{-2} \;\text{G})$ and $t_{100} = t/(100\;\textrm{yr})$. Thus, for these systems the cooling break is expected to lie in the sub-millimeter band. A younger age would shift this cooling break to higher frequencies.

The self-absorption frequency $\nu_\mathrm{sa}$ depends weakly on the energy content of the nebula and is given by
\begin{equation}
  \nu_{\rm sa}  \approx  230 \;  R_{17.3}^{-\frac{6}{5}} \; B_{-2}^{\frac{3}{10}}\; \nu_{9.3}^{-\frac{1}{2}} \; L_{29.3}^{\frac{3}{5}} \;\text{MHz}\;,    
\end{equation}
which shows the self-absorption frequency is strongly constrained and decided mainly by the MWN's radius.

In summary,  \textit{the observation of a cooling break in the PRS spectra in the sub-millimeter band can validate the optimal magnetar parameter space detailed above.}

\subsection{Implications for the \texorpdfstring{$ \alpha-\Omega $}{alpha-Omega} dynamo}

Our optimal parameter space requires an extremely strong internal magnetic field of $\sim 10^{16.5}\;$G with a longer initial rotation period of $P_{\rm i} \geq 10\;$ms. These conditions are inconsistent with the assumptions of the $\alpha-\Omega$ dynamo model proposed by \cite{Duncan-Thomson92}, which requires an initial period of $P_{\rm i} \lesssim 3\;$ms
(convective time) for the dynamo to amplify the seed magnetic field to $10^{15} - 10^{16}\;$G. Instead, our findings favor the alternative flux-freezing model proposed by \cite{Ferrario06} as a viable explanation. This is further supported by observational studies of supernova remnants (SNRs) surrounding magnetars \citep{Vink06}, which indicate SNR energetics comparable to those expected in normal supernovae. In contrast, the dynamo model predicts significantly enhanced nebular expansion and kinetic energy due to the rapid loss of rotational energy from a millisecond magnetar.

\subsection{Observational implication of rotation-powered phase of ultra-strong magnetars}\label{obs_feas} 

Here we explore the feasibility of observing a luminous PRS ($L_{\nu_{\rm m}} \geq 10^{29}\;$erg\;s$^{-1}$\,Hz$^{-1}$) at GHz frequencies predominantly in the rotation-powered phase for ultra-strong magnetars with $B_{\rm int,16} =3$ \textit{for any class of supernova explosion}. For simplicity, we assume an initial spin-period of $P_{\rm i} = 20\;$ms. For this choice, the nebular expansion is predominantly due to the energy imparted during the supernova explosion, $E_\mathrm{SN}$. 

Table \ref{tab:SN_type} summarizes the observables in the rotation-power dominated space for various cases of the supernova explosion energy and the corresponding mass of the ejecta. We find that in the rotation-dominated phase irrespective of $E_\mathrm{SN}$ the nebula is too compact where self-absorption effects make it unlikely for any significant radio emission to escape. Further, the DM at these ages is too large, even assuming an ionization fraction of just 1 percent assuming shock heating (see Eq.~16 of \citealt{PG2018}  which suggests $f_{\rm ion} = 10^{-2}$ from shock heating alone). This also makes it highly implausible for FRB signals to be detected on such timescales. 

For the minimum internal magnetic field strength required for FRB production (see \S \ref{req_FRB}), with $B_{\rm int,15} = 1$, the corresponding transition timescale from Eq.~(\ref{time_switch}) is $t_{\rm switch}\approx3.4\;$kyr, which exceeds the Sedov-Taylor timescale (Eq.~(\ref{time_ST})). This occurs because, for $\alpha_{\rm B} =1$, the internal magnetic field decay timescale from Eq.~(\ref{mag_param}) is $t_{\rm d} \approx 10\;$kyr. Consequently, most of the internal magnetic energy is released too late, when the nebula has expanded significantly and the corresponding low energy density is insufficient to power the high radio luminosity of the PRS. At this FRB-limiting internal magnetic field strength, the rotation-powered phase persists throughout the free-coasting phase and beyond. However, to produce a compact and luminous PRS, the rotational energy must be extracted early, near the spin-down timescale $t_{\rm sd} \approx 3\times 10^6 \; P_{\rm i,-2}^2 B_{\rm d,14}^{-2}\;$s, when most of $E_{\rm rot}$ is released. This is indeed the case for the candidate low-luminosity PRS associated with 20201124A with a $B_\mathrm{d,14} = 1$ and $P_\mathrm{i,-2}=1$ can only if the current age is $t<20$ years.

\begin{table*}
\centering
\caption{Rotation-powered parameter space for $B_{\rm int,16} = 3$ and initial spin period $P_{\rm i} = 20$ ms.  For the (theoretical) maximum internal magnetic field strength $B_{\rm int,16} = 3$ and assuming a internal magnetic field decay-power index $\alpha_{\rm B} = 1$ , the rotation-powered phase lasts $t_{\rm switch} \approx 0.7$ year (less than a year). The ionization fraction in the supernova ejecta has been assumed to be $f_{\rm ion} = 0.01$. }    
\begin{tabular}{|c|c|c|c|c|c|c|} \hline 
      Scenario   & ($E_{\rm SN}, M$) & $v_{\rm SNR}$ & $t_{\tau}$ & $R_{\rm switch}$ & $\nu_{\rm sa} (R_{\rm switch})$ & DM $(R_{\rm switch})$ \\
                 & (erg, $M_{\odot}$ ) & ($10^3\,$km/s) & (yr) & (pc) & (GHz) & (pc cm$^{-3}$) \\   \hline
      Superluminous SN   & 10$^{52}$ erg, 3 $M_{\odot}$ \;   & 18.3 & $10^{-1}$ & $10^{-2}$ & $>6$ & $5.2 \times 10^{2}$ \\ 
       Core-collapse SN   &  10$^{51}$ erg, 3 $M_{\odot}$   & 5.8  & $4 \times 10^{-1}$  & $4 \times 10^{-3}$ & $>25$ & $5.2 \times 10^3$  \\
        Ultra-stripped SN   & 10$^{50}$ erg, 0.1 $M_{\odot}$   & 10.1   & $4 \times 10^{-2}$ & $7 \times 10^{-3}$  & $>13$ & $5.8 \times 10^1$ \\ 
      Sub-energetic SN   & 10$^{50}$ erg, 10 $M_{\odot}$   & 1 & 4.3 & $7 \times 10^{-4}$  & $>207$ & $5.8 \times 10^{5}$   \\  \hline 
\end{tabular} \label{tab:SN_type}
\end{table*}

\subsection{Speculations on the correlated RM-DM variation}\label{RM_DM_correlation}
Here we briefly comment on the RM-DM variation \citep{Hilmarsson2021} in magnetar powered MWN scenario. We modeled baryonic outflows from giant flares (GFs) as a continuous magnetic luminosity. In practice, these outflows occur episodically as ion-electron GFs. The proper speed $\Gamma\beta$ of a GF can vary widely, ranging from $\sim$1 to $\sim$100, while the GF isotropic equivalent energy output spans $E_{\rm GF} \sim 10^{44} - 10^{46}\;$erg \citep{2006ApJ...638..391G,2024arXiv241116846B}. For magnetars within a MWN, after a GF outflow is ejected it decelerates when it collides with the shocked wind behind the termination shock, subsequently reaching pressure equilibrium with the hot electron-positron gas expelled through dipolar spin-down. Since these GF outflows contain unpaired electrons, the  rotation measure (RM) predominantly arises from the unpaired non-relativistic electrons (see Equation \ref{RM}). If the same electrons also contribute to the source's dispersion measure (DM), this naturally explains the correlated RM-DM variations observed in these sources. The RM is determined by the number of GF blobs intersecting the line of sight, while the motion of the blobs is governed by the thermal pressure of the hot MWN and turbulence in the medium (not accounted for in our one-zone model). Due to significant uncertainties in blob motion, modeling the temporal variation of RM lies beyond the scope of this work. Nonetheless, as the sound speed in the MWN is $\approx c/\sqrt{3}$, sub-sonic turbulence can potentially still lead to fast large amplitude RM variations.

In what follows, we provide some fiducial estimates of the RM contribution. The contribution to RM is primarily due to unpaired non-relativistic (thermal) electrons in the wind nebula and can be expressed as:
\begin{equation}
{\rm RM} \approx \frac{e^3}{2\pi m_e^2 c^4} \frac{\chi N_{\rm nr} \langle B_{\parallel} \rangle R_n}{(4/3)\pi R_n^3} \approx \frac{\chi}{\zeta_e} \frac{1700}{R_{17}^2} \; L_{\nu,29} \; \text{rad} \; \text{m}^{-2}  \label{RM}
\end{equation}
where the non-relativistic electrons in the MWN is represented in terms of the relativistic electrons $N_{\rm r}$ though the ratio $\zeta_{\rm e}$ through $N_{\rm nr} = \frac{N_{\rm r}}{\zeta_{\rm e}} $, $\langle B_{\parallel} \rangle$ is the average component of the magnetic field along the line of sight and $\chi$ represents the ratio of unpaired non-relativistic electrons to  the total non-relativistic electrons. Equation \ref{RM} shows that in the one-zone framework, RM scales linearly with the source spectral luminosity (in agreement with the trend in \citealt{Yang20}).

\subsection{Summary and observational prospects}

In summary, the properties of the PRSs linked to FRBs 20121112A and 20190520B suggest that they are powered by a magnetar with an internal magnetic field of $\sim 10^{16} - 10^{16.5}\;$G with decay timescale $t_{\rm d} \sim 10 -10^{2.5}$, a dipolar magnetic field of $\sim 10^{15}-10^{15.5}\;$G, and a moderate spin period ($P \gtrsim 10$ ms), embedded in ejecta ($M \gtrsim 3-10 M_{\odot}$). The sub-energetic supernovae with massive ejecta favor the slower decay timescales of the internal magnetic field. Such conditions are rare, suggesting only a small subset of magnetars produce these sources.

\textit{We recommend observing the PRS of FRB 20121102A and FRB 20190520B to measure the self-absorption frequency ($\sim$ 200 MHz) at a flux density of $\sim$ 180 $\mu$Jy and cooling break ($\sim $ 150 -200 GHz) at a flux density of $\sim$ 40-20 $\mu$Jy and using high-resolution imaging to confirm a PRS size of $\sim$ 0.1 pc. Failure to detect these features would strongly disfavor a magnetar model. We also recommend searching for an SED turnover in the candidate PRS of FRB 20201124A; if absent by $\sim$ 150 GHz, the rotationally powered MWN scenario can be ruled out with high confidence. We also encourage long-term follow-up of the flux of the PRS at a given band to provide independent constraints on the ages of these PRSs (see appendix \ref{ap:flux_dec} for illustration for QRS121102).}

Further observations  of new PRS candidates \citep{Bruni24,2024arXiv241213121B} at cm and sub-mm are encouraged to constrain nebular energy and age. . The implications  for magnetar-based models for  non-detection of PRSs from  FRB repeaters will be pursued in a separate study.  

\begin{acknowledgments}
PB is supported by a grant (no. 2020747) from the United States-Israel Binational Science Foundation (BSF), Jerusalem, Israel, by a grant (no. 1649/23) from the Israel Science Foundation and by a grant (no. 80NSSC 24K0770) from the NASA Astrophysics Theory Program. 
\end{acknowledgments}

\bibliography{References}{}
\bibliographystyle{aasjournal}

\appendix

\section{Flux variation for QRS121102- A diagnostic of the current age}\label{ap:flux_dec}

Here we explore how the flux (at $\nu_\mathrm{obs}=1.5\;$GHz) of the PRS associated with FRB121102 (a.k.a. QRS121102) changes with its age. 

We assume the following model for the radio flux density at a given radio frequency $\nu_\mathrm{obs}$ as
\begin{equation}
     F_\nu(T_\mathrm{obs}) = K (T_\mathrm{obs}- T_\mathrm{birth})^{-\alpha_\mathrm{f}}\;,  \label{fit_flux}
\end{equation}
where $K$ is a normalization constant, $\alpha_{\rm f}$ is the flux decay index, $T_\mathrm{obs}$ is the absolute time for the observation and $T_\mathrm{birth}$ is the time of birth of the system. For comparison at the same frequency, we use the flux measurement in the VLA L band using observations reported in \cite{Chatterjee2017} ($T_\mathrm{obs,1}$) and \cite{Yang2024} ($T_\mathrm{obs,2}$) spaced $\sim$ 7 years apart. The current age $t \equiv T_\mathrm{obs,2} - T_\mathrm{birth}$ is taken as the difference $T_{\rm obs,2} - T_{\rm birth}$. 

\begin{figure*}
\begin{tabular}{cc}
 \includegraphics[scale=0.45]{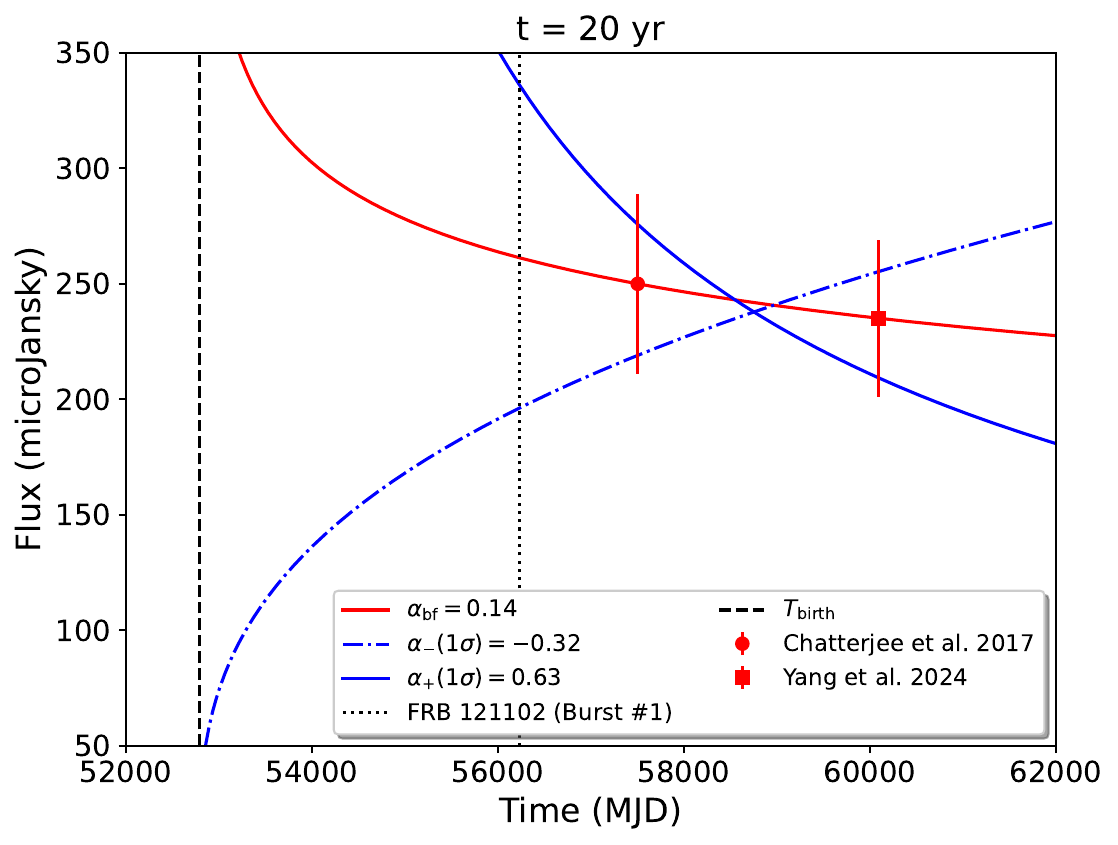} & \includegraphics[scale=0.45]{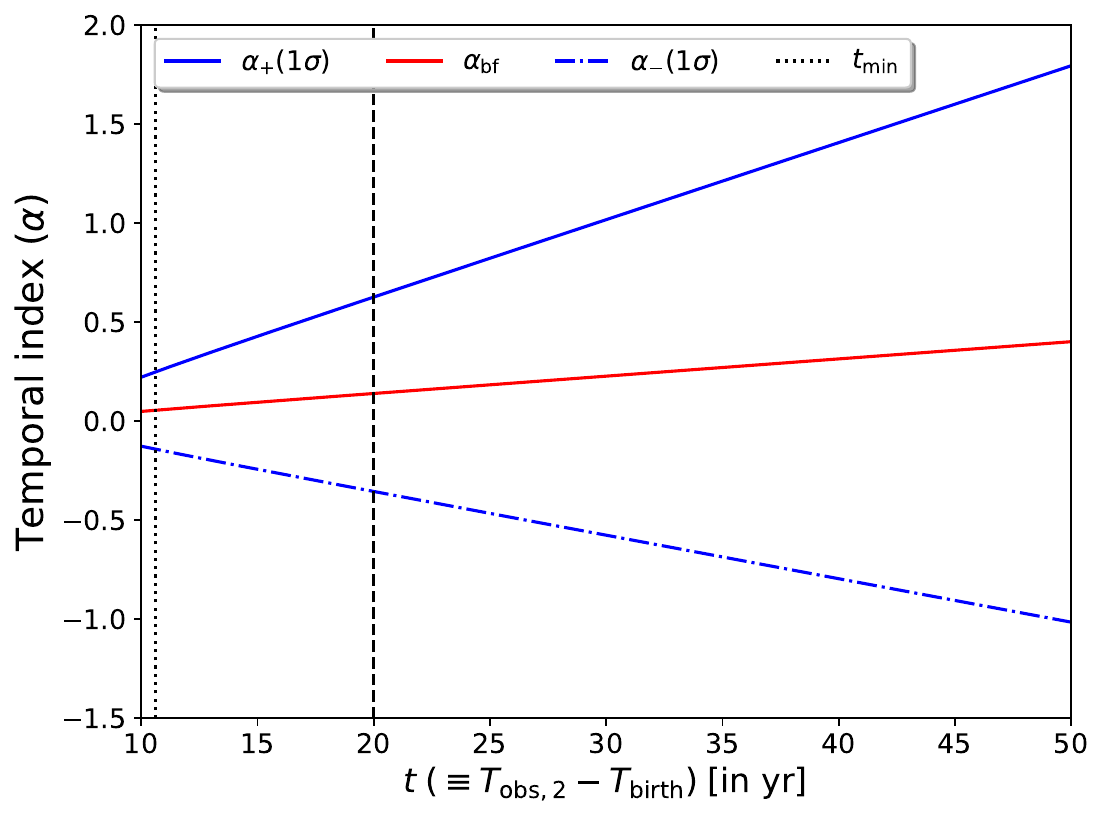}    \\
\end{tabular}
\caption{$\chi^2$ fit to the flux variation of QRS121102 at L-band of VLA. \textbf{Left:} shows the best fit (in red) \ref{fit_flux} passing through the mean values of the flux $250 \pm 39 \,\mu\text{Jy}$ at $T_\mathrm{obs,1} = 57504.1$ (shown as a red circle from \citealt{Chatterjee2017}) and the flux $235.1 \pm 39 \,\mu\text{Jy}$ at $T_\mathrm{obs,2} = 60092.6$ MJD (shown as a red square from \citealt{Yang2024}). The vertical dotted black line corresponds to the observation date $T_\mathrm{FRB,\#1} = 56233.3$ of the first FRB burst from FRB 121102 (\citealt{spitler2014}). The solid blue and dot-dashed line show the $1\sigma$ deviation from the best fit with $\alpha_{+} > \alpha_\mathrm{bf}$ and $\alpha_{+} < \alpha_\mathrm{bf}$, respectively. 
\textbf{Right:} shows the fits as a function of the current age $t \equiv T_{\mathrm{obs,2}} - T_{\mathrm{birth}}$, where the observation date reported in \cite{Yang2024} has been taken as the reference time of observation. The vertical dotted black line shows the minimum age of the PRS $t_\mathrm{min} \equiv T_\mathrm{obs,2}-T_\mathrm{FRB,\#1} \sim 10.6$ years.\label{chisqr_fits}}
\end{figure*}

Figure \ref{chisqr_fits} shows the $\chi^2$ fit to the two data points mentioned above. It can be seen that for younger ages of the system, the best-fit flux decay index $\alpha_\mathrm{bf}$ is shallow and vice versa. This shows that long-term monitoring of the PRSs can shed light on the age of these systems.

\section{Termination shock radius in the free-coasting phase}\label{app:term-rad}

The termination shock radius $R_{\rm TS}$ can be obtained by equating the ram-pressure of the outflowing cold MHD wind $p_{\rm ram} = \frac{\dot{E}}{4 \pi R_{\rm TS}^2 c}$  to the thermal pressure of the hot MWN $p_{\rm neb} = (\hat{\gamma} - 1) e_{\rm neb} = \frac{1}{3} e_{\rm neb} =  \frac{\dot{E} t}{4 \pi R_{\rm neb}^3 }$ (where we assume $\hat{\gamma} =\frac{4}{3}$ for the hot MWN ). This can be expressed as 
\begin{equation}
   \frac{\dot{E}}{4 \pi R_{\rm TS}^2 c} = \frac{\dot{E} t}{4 \pi  (v_{\rm SNR} t)^3}  \Rightarrow \frac{R_{\rm TS}}{R_{\rm neb}} = \sqrt{\frac{v_{\rm SNR}}{c}} \hspace{0.5 cm} \text{for $t_{\rm sd}<t< t_{\rm dec}$}\;, 
\end{equation}
where we used the relation $R_{\rm neb} = v_{\rm SNR} t$. 

\section{Contribution of synchrotron flux from the forward shocked ISM}\label{app:SNR_flux}

We estimate here the synchrotron flux contribution from the forward shock being driven into the ISM by SNR. Let us consider a strong Newtonian forward shock being driven into the ISM by a supernova remnant. In the rest frame of the forward shock front the quantities in the unshocked upstream and the shocked downstream can be labeled by 1 and 2 respectively. In the strong shock-regime the adiabatic constant downstream can be taken as $\hat{\gamma} = \frac{5}{3}$ while the particle density is four times the particle density in the unshocked ISM ($n_{2} = 4 n_{1}$). The conservation of mass $(n_1 v_1 = n_2 v_2)$ across the shock front gives the relative velocity of the upstream to downstream as $v_{\rm ud} \approx \frac{3}{4} v_{\rm SNR}$.  This gives the internal energy in the downstream region as
\begin{equation}
    e_{\rm 2,int} = 2 n m_{\rm p} v_{\rm SNR}^2 \approx 1.1 \times 10^{-6} \; n_{\rm o}^{\frac{1}{2}} \; E_{\rm SNR,51} \; M_{3}^{-1} \; \text{erg cm$^{-3}$},
\end{equation}
where $n_{\rm o} = n_1/ 1\; \text{cm$^{-3}$}$ is the particle density in the unshocked ISM. Assuming a fraction $\epsilon_{\rm B}$ of the internal energy density is channeled into post-shock magnetic field in the downstream can be estimated as 
\begin{equation}
\begin{split}
    B_{2} &\ = \sqrt{8 \pi \epsilon_{\rm B} e_{\rm 2,int}}  \approx 0.5 \; \epsilon_{\rm B,-2}^{\frac{1}{2}} n_{\rm o}^{\frac{1}{2}} \;  E_{\rm SNR,51}^{\frac{1}{2}} \; M_{3}^{-\frac{1}{2}}  \; \text{mG},\label{B_shock}
\end{split}
\end{equation}
where $\epsilon_{\rm B,-2} = \epsilon_{\rm B}/10^{-2}$.

Consider a nonthermal distribution of electrons due to shock acceleration across the forward shock with power-law index $p>2.5$. For synchrotron emission the average LF of the non-thermal particles needs to be at least mildly relativistic. This is difficult as in the Newtonian regime, the available internal energy per baryonic particle is much less than the rest mass of the particle, i.e. $\frac{e_{\rm2,int}}{\rho_2 c^2} \approx 10^{-9}\;  \ll 1$. This situation can be circumvented by requiring that a fraction $\epsilon_{\rm e}$ of the internal energy density $(=\epsilon_{\rm e} \; e_{\rm int,2})$ is channeled into a very small fraction $\xi_{\rm e} \ll 1$ of shock-accelerated nonthermal electrons. In such a deep Newtonian scenario \citep{Granot06,Sironi13,Beniamini22}, the minimal LF of the electrons can be mildly relativistic $\gamma_{\rm m} = \gamma_{\rm dn} = \sqrt{2}$ (a fiducial value). The corresponding minimal synchrotron frequency can be estimated as,
\begin{equation}
    \nu_{\rm m} \equiv \gamma_{\rm dn}^2 \nu_{\rm B2} \approx 2.8 \;  \epsilon_{\rm B,-2}^{\frac{1}{2}} n_{\rm o}^{\frac{1}{2}} \;  E_{\rm SNR,51}^{\frac{1}{2}} \; M_{3}^{-\frac{1}{2}}  \;  \text{kHz},
\end{equation}
where $\nu_{\rm B2} \approx 1.4 $ kHz is the cyclotron frequency corresponding to the post-shock magnetic field evaluated in equation \ref{B_shock}. It must be mentioned here that the minimal frequency is less than the plasma frequency $\nu_{\rm p} \approx 8.9 \; n_{\rm o}^{1/2}$ kHz of the shocked medium. The minimal LF is given by $\gamma_{\rm m} = \gamma_{\rm dn} = \frac{p-2}{p-1} \frac{\epsilon_{\rm e}}{\xi_{\rm e}} \frac{m_{\rm p}}{m_{\rm e}} \frac{v_{\rm SNR}^2}{2c^2} $ which can be inverted to get 
\begin{equation}
    \xi_{\rm e} \approx 8 \times 10^{-3} \; \frac{g(p)}{g(2.5)} \epsilon_{\rm e,-1} \; \; E_{\rm SNR,51}^{2} \; M_{3}^{-2},  
\end{equation}
where $g(p) = \frac{p-2}{p-1} $ with $g(2.5) = \frac{1}{3}$ and $\epsilon_{\rm e,-1} = \epsilon_{\rm e}/0.1$. The number of non-thermal electrons contributing to the emission is $N_{\rm e} \approx \frac{4 \pi}{3} \pi R^3 n  \xi_{\rm e}$ with each non-thermal electron contributing $P_{\nu} \approx \frac{m_{\rm e} c^2}{3 q_{\rm e} } \sigma_{\rm T} B_2 $. 
The spectral luminosity $L_{\nu_{\rm m}}$ at $\nu_{\rm m}$  can be estimated as 
\begin{equation}
    L_{\nu_{\rm m}} \approx 4.2 \times 10^{25}  \; \frac{g(p)}{g(2.5)} \epsilon_{\rm e,-1} \; \; \epsilon_{\rm B,-2}^{\frac{1}{2}} \; n_{\rm o}^{\frac{1}{2}} \; t_{10}^3\; 
     E_{\rm SNR,51}^{\frac{3}{2}} \; M_{3}^{-\frac{3}{2}}\; \text{erg s$^{-1}$ Hz$^{-1}$},
\end{equation}
where $t_{10} = t/10\; \text{yr}$. At $\nu_{\rm GHz} = 1 $ GHz, the expected spectral luminosity (assuming p =2.5) is 
\begin{equation}
     L_{\nu_{\rm GHz}} = L_{\nu_{\rm m}} \left( \frac{\nu_{\rm GHz}}{\nu_{\rm m}} \right)^{-\frac{(p-1)}{2}}  \approx 3 \times 10^{21} \; \frac{g(p)}{g(2.5)} \epsilon_{\rm e,-1} \; \; \epsilon_{\rm B,-2}^{\frac{7}{8}} \; n_{\rm o}^{\frac{7}{8}} \; t_{10}^3\; 
     E_{\rm SNR,51}^{\frac{15}{8}} \; M_{3}^{-\frac{15}{8}}\;   \text{erg s$^{-1}$ Hz$^{-1}$} 
     \ll 10^{29} \;  \text{erg s$^{-1}$ Hz$^{-1}$} \label{LGHz}
\end{equation}

Equation \ref{LGHz} shows the forward shocked ISM has negligible contribution to the observed PRS emission at GHz frequencies. This leaves MWN as the only promising source of powering PRS. This is also expected on physical grounds as the non-thermal electrons in the forward shocked ISM are at most mildly relativistic while in MWN the charged particles are ultra-relativistic electron-positron pairs.
In \S \ref{spec_radio} we examine how the MWN energy content and size (radius) of the nebula influence key observables, such as the peak flux and peak frequency of the radio spectrum.

\section{Estimation of $B_\mathrm{d,max,rot}$} \label{max_rot_Bd}

\begin{figure}
    \centering
    \includegraphics[scale=0.55]{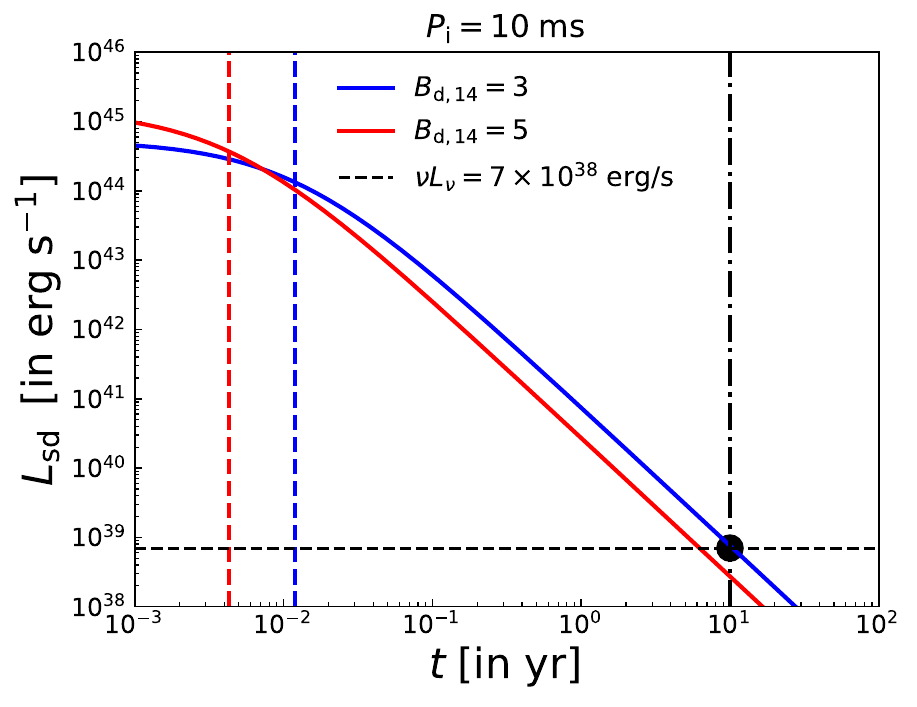}
    \caption{Illustration of the dipolar magnetic field strength, that for a fixed initial spin period $P_{\rm i} = 10$ ms, can power a luminous source with radiative output $\nu L_{\nu} = 7 \times 10^{38}$ erg s$^{-1}$ (horizontal black dashed line) at time $t=$ 10 years (vertical black dot dashed line) through spin-down due to magnetic braking $(m=2)$ being efficiently channeled to non-thermal electrons with $\epsilon_\mathrm{e} = 1$. The blue solid line corresponds to $B_\mathrm{d,max,rot} = 3 \times 10^{14}$ gauss (see equation \ref{Bdmax}) that can sustain $\nu L_{\nu}$ at time $t$ (represented as a black point). The red solid line corresponds to $B_{\rm d} = \frac{5}{3} B_{\rm d,max,rot} > B_{\rm d,max,rot}$. Evidently, it lies below the black point meaning it cannot account for the observed radiative output.    }
    \label{fig:Max_Bd}
\end{figure}

In this section,  we aim to estimate the the maximum dipolar magnetic field $B_\mathrm{d,max,rot}$ required to sustain a PRS with luminosity $\nu L_{\nu}$ at age $t$ in the rotational loss channel. The rotational channel quenches for all $B_{\rm d} > B_{\rm d,max,rot}$.

The maximum dipolar field strength $B_{\rm d,max,rot}$ for sustaining the rotational channel can be obtained by equating the energy injection rate into the non-thermal electrons $\epsilon$ at time $t$ to the luminosity $\nu L_{\nu}$ of the PRS as (for $t \gg t_{\rm sd}$), 
\begin{equation}
   \epsilon_{\rm e} L_{\rm o} \left( \frac{t}{t_{\rm o}}\right)^{-m} = \nu L_{\nu} \Rightarrow B_{\rm d,max,rot} = \sqrt{\frac{I c^3 }{2 f \Omega_{\rm i}^2 R_{\rm NS}^6}} \left[ \frac{\nu L_{\nu} t^{\rm m}}{\epsilon_{\rm e} E_{\rm rot}} \right]^{\frac{1}{2(1-m)}},
\end{equation}
which for magnetic dipole breaking ($m=2$) is given as 
\begin{equation}
 B_{\rm d,max,rot}  = \sqrt{\frac{\epsilon_{\rm e} I^2 c^3}{4 f R_{\rm NS}^6}} \frac{1}{\nu L_{\nu}} \frac{1}{t}   \hspace{1cm}  \text{for m=2 (magnetic dipole breaking)}. \label{Bdmax} 
\end{equation}

While the derivation above always holds for fast cooling electrons that must have been injected in the last dynamical time. For slow cooling electrons as is required for our scenario, they can in general be dominated by injection at earlier times (meaning the current nebular energy content) despite adiabatic cooling. However, in our set-up, for times $t$ far-away from the characteristic timescales $t_\mathrm{sd}$ and $t_\mathrm{d,0}$,  the current nebular energy is comparable to the energy injected in the last dynamical timescale and thus the derivation is still valid. 

Figure \ref{fig:Max_Bd} shows that for a given $\nu L_{\nu}$ at time $t$ a higher dipolar magnetic field ($B_\mathrm{d} > B_{\rm d,max,rot}$) shortens the spin-down timescale and reduces rotational energy release at time $t$.

\section{Switching from rotation dominated to  magnetic dominated channel}
\label{app:rot_mag_switch}

\begin{figure}
    \centering
    \includegraphics[scale=0.55]{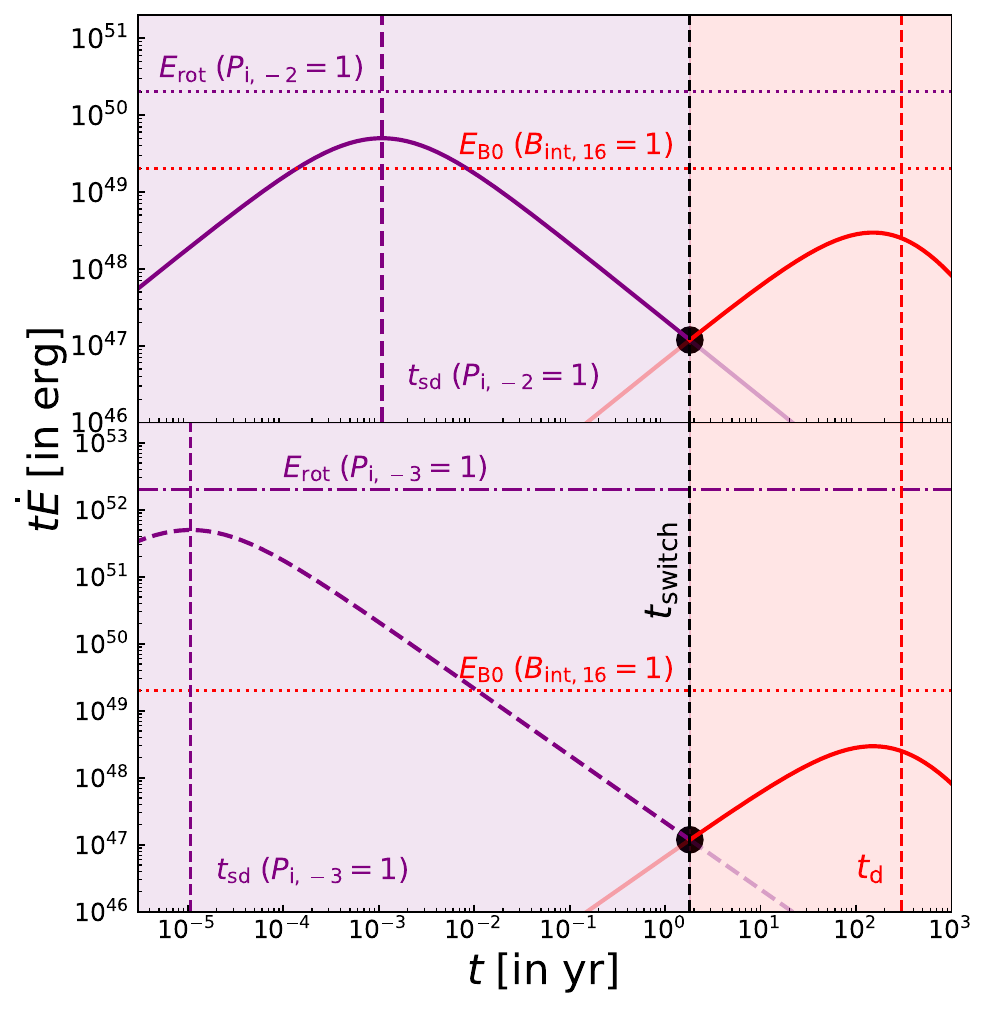}
    \caption{Illustration of nebular energy content switching from rotation dominated channel (shown in shaded purple)  to internal magnetic field decay dominated channel (shown in shaded red). The parameters assumed in both panels are $m =2,\; t_{\rm d} = 300\; \text{yr},\; \epsilon_{\rm out} =1,\; B_{\rm int} = 10^{16}\; \text{G}, B_{\rm d} = 10^{14}\; \text{G}$.  The top panel corresponds to an initial spin period of 10 milliseconds while the bottom panel corresponds to an initial spin period of 1 milliseconds. \textit{Top panel:} the horizontal dotted purple and  lines corresponds to the initial rotational kinetic energy reservoir $E_{\rm rot} (P_{\rm i} = 10 \;\text{ms})$ and the initial internal magnetic field energy reservoir $E_{\rm B,0}(B_{\rm int} = 10^{16}\; \text{G})$ respectively. The vertical thicker dashed purple line corresponds to the spin-down timescale $t_{\rm sd}(P_{\rm i,-2},B_{\rm d,14})$. \textit{Bottom panel:} the horizontal dot-dashed purple and  lines corresponds to the initial rotational kinetic energy reservoir $E_{\rm rot}(P_{\rm i} = 1\;\text{ms})$ and the initial internal magnetic field energy reservoir $E_{\rm B,0}(B_{\rm int} = 10^{16}\; \text{G})$ respectively. The vertical thinner dashed purple line corresponds to the spin-down timescale $t_{\rm sd}(P_{\rm i,-3},B_{\rm d,14})$.  In both panels the vertical dashed black line and the vertical dashed red line corresponds to the switching timescale and the spin-down timescale respectively. The black dot in both panels denotes the half of the total energy content in the nebula at $t_{\rm switch}$. It can be seen that the location of this point does not depend on the initial spin period (see text for detailed explanation).  }   \label{fig:switch_time_illustration}
\end{figure}

In this section, we aim to study the dynamics of energy injection into the nebula through both the rotation and the internal magnetic field decay channel.

For energy loss rate $\dot{E}$ through any generic channel energy, injected energy over a dynamical timescale $t$ is $t \dot{E}$. Figure \ref{fig:switch_time_illustration}
shows how for both the spin-down rotational loss (shown in purple) and the internal magnetic decay (shown in red), the energy injected over each dynamical timescale first increases, reaches maximum and then decreases. 
 The maximum energy injected over any dynamical timescale is at a characteristic timescale $t_{\rm ch}$, which can be estimated from $\frac{d}{dt} (\dot{E}t)\lvert_{t=t_{\rm ch}} = 0$ with the maximum injected energy over one dynamical timescale being $E_{\rm max,dyn,inj} = t_{\rm ch} \dot{E}(t_{\rm ch})$. For the rotation loss and the internal magnetic field decay, the characteristic timescale is the spin-down timescale $t_{\rm sd}$ and the internal magnetic field timescale $t_\mathrm{d}$, respectively.  The maximum injected energy at $t=(t_{\rm sd},t_{\rm d})$ is given as,
\begin{equation}
    E_{\rm max,dyn,inj} = 
    \begin{cases}
        &\  \frac{E_{\rm rot}}{2^{\rm m}} = \frac{E_{\rm rot}}{4} = 5 \times 10^{51} \;P_{\rm i,-3}^{-2} \; \text{erg} \hspace{0.4cm}  \text{for $m=2$},       \\
        &\  \frac{\alpha_{\rm B} E_{\rm B,0}}{2^{\frac{2}{\alpha_{\rm B}} +1 }  } = \frac{E_{\rm B,0}}{8}  = 2.5 \times 10^{48} B_{\rm int,16} \; \text{erg} \hspace{0.4cm} \text{for $\alpha_{\rm B} = 1$}.
    \end{cases} \label{max_enj}
\end{equation}

Next, we aim to estimate the time $t_{\rm switch}$ when the energy content in the nebula switches from the rotation-powered channel to the internal magnetic field decay powered channel. We are interested in the limit when the switching time follows the following ordering $t_{\rm sd} \ll t_{\rm switch} \ll t_{\rm d}$. Under this approximation, the time of switching can be obtained by finding the intersection of the power-law approximations of the rotation and internal magnetic field decay energy curves. The first one is the power law approximation to the decaying for the rotational energy curve given as $\frac{E_{\rm rot }}{t/t_{\rm sd}}$ while the second one is the power-law approximation $\frac{\epsilon_{\rm out}\; E_{\rm B,0} t}{t_{\rm d}}$ for the rising internal magnetic field decay curve.  This can be represented as 
\begin{equation}
    \frac{E_{\rm rot} t_{\rm sd} }{t_{\rm switch}} = \frac{\epsilon_{\rm out} E_{\rm B,0} t_{\rm switch}}{t_{\rm d}} \Rightarrow t_{\rm switch} = \sqrt{\frac{E_{\rm rot} t_{\rm d} t_{\rm sd}}{\epsilon_{\rm out} E_{\rm B,0}}} = \sqrt{\frac{3 I^2 c^3}{2 f R_{\rm NS}^{9}}} \frac{1}{\epsilon_{\rm out}^{\frac{1}{2}} B_{\rm int} B_{\rm dip}} \sqrt{t_{\rm d}}, 
\end{equation}
which can be simplified using $\frac{B_{\rm dip}}{B_{\rm int}} = \sqrt{f_{\rm dip}}$ (following \citealt{Beniamini25}) and following \citealt{Colpi2000} we have $t_{\rm d} \propto B_{\rm int}^{-\alpha_{\rm B}}$ as 
\begin{equation}
    t_{\rm switch}  = \sqrt{\frac{3 I^2 c^3}{2 f R_{\rm NS}^{9}}} \frac{1}{\epsilon_{\rm out}^{\frac{1}{2}}} \frac{t_{\rm d}^{\frac{1}{2}}}{f_{\rm dip}^{\frac{1}{2}}} B_{\rm int}^{-2 - \frac{\alpha_{\rm B}}{2}}. \label{switch}
\end{equation}
such that the nebular energy content and the free coasting radius of the supernova remnant (and the nebula) at $t_{\rm switch}$ can be obtained as,
\begin{eqnarray}
    &\ E_{\rm neb} (t_{\rm switch}) =  \frac{2 E_{\rm B,0} t_{\rm switch}}{t_{\rm d}} = \sqrt{\frac{4 I^2 c^2}{3 f R_{\rm NS}^5}} \frac{B_{\rm int}^{1+ \frac{\alpha_{\rm B}}{2}}}{\epsilon_{\rm out}^{\frac{1}{2}} f_{\rm dip}^{\frac{1}{2}} t_{\rm d}^{\frac{1}{2}}}                 \\
    &\ R(t_{\rm switch}) = v_{\rm SNR} t_{\rm switch} = \sqrt{\frac{3 I^2 c^3}{ f R_{\rm NS}^{9}}} \frac{1}{\epsilon_{\rm out}^{\frac{1}{2}}} \frac{t_{\rm d}^{\frac{1}{2}}}{f_{\rm dip}^{\frac{1}{2}}} B_{\rm int}^{-2 - \frac{\alpha_{\rm B}}{2}} \; E_{\rm SNR}^{\frac{1}{2}} \; M_{\rm SNR}^{-\frac{1}{2}} .
\end{eqnarray}

Equation (\ref{switch}) shows that the switching time is independent of the initial rotation period $P_{\rm i}$ of the magnetar. This can also be seen from figure \ref{fig:switch_time_illustration}. This is because for times much larger than the spin-down timescale the rotational energy loss becomes independent of $P_{\rm i}$.

\section{Inferring Nebular energy from observables}\label{ap:neb_en}

We assume a non-thermal distribution as represented in equation \ref{non-thermal}. The average Lorentz factor of the particles is
\begin{equation}
    \langle \gamma \rangle_\mathrm{e} \approx \frac{p-2}{p-1} \gamma_{\rm m} \hspace{2cm} \text{for $p>2$}. \label{avg_LF}
\end{equation} 

The spectral luminosity at $L_{\nu_{\rm m}}$ is given as
\begin{equation}
    L_{\nu_{\rm m}} = \int^{\infty}_{\gamma_{\rm m}}\; d\gamma_\mathrm{e}  \; \frac{dN}{d \gamma_\mathrm{e}} \; P_{\rm \nu,max}  \left( \frac{\nu_\mathrm{m}}{\nu} \right)^{\frac{1}{3}} = \left( \frac{p-1}{3p-1}\right) \frac{\sigma_\mathrm{T} B}{q_\mathrm{e}} N_{\rm rel} m_\mathrm{e} c^2. \label{Lum_max} 
\end{equation}
where the upper limit for $\gamma_\mathrm{e}$  is taken to be $\infty$ as for $p>2$ both the total energy and number of non-thermal electrons is dominated by particles closer to $\gamma_\mathrm{m}$. We used the identity $\frac{\nu_\mathrm{m}}{\nu} = \frac{\gamma^2_\mathrm{m}}{\gamma^2}$ and $P_\mathrm{\nu,max} =  \frac{m_\mathrm{e} c^2}{3 q_\mathrm{e}} \sigma_\mathrm{T} B$

Equation \ref{Lum_max} can be inverted to get the total number of non-thermal electrons in the nebula given as,
\begin{equation}
    N_{\rm rel} m_{\rm e} c^2 = \left( \frac{3p-1}{p-2} \right) \frac{q_{\rm e}}{\sigma_{\rm T} B} L_{\nu_{\rm m}}
\end{equation}

The total energy carried by the non-thermal electrons are given as
\begin{equation}
    E_{\rm rel} = \left( \langle \gamma \rangle_\mathrm{e} - 1 \right) N_{\rm rel} m_{\rm e} c^2 \approx \langle \gamma \rangle_\mathrm{e} N_{\rm rel} m_{\rm e} c^2 = \left[  \left( \frac{2 \pi m_{\rm e} c}{q_{\rm e}}\right)^{1/2}  \frac{q_{\rm e}}{\sigma_{\rm T}} \right] \left( \frac{3p-1}{p-2} \right) \frac{\nu_{\rm m}^{1/2} L_{\nu_{\rm m}}}{B^{3/2}}. \label{En_rel} 
\end{equation}

The observables are peak spectral luminosity $L_{\nu_\mathrm{m}}$ and peak frequency $\nu_\mathrm{m}$.

\section{Inferring nebular dynamics from the PRS spectrum}\label{ap:neb_dyn}

Using equation \ref{En_rel} the average magnetic field $B$ in the nebula can be estimated as 
\begin{equation}
B =     \left[  \left( \frac{2 \pi m_{\rm e} c}{q_{\rm e}}\right)^{1/3}  \left( \frac{q_{\rm e}}{\sigma_{\rm T}} \right)^{2/3} \right] \left( \frac{3p-1}{p-2} \right)^{2/3} \; \nu_{\rm m}^{1/3} \; L_{\nu_{\rm m}}^{2/3} \; E_{\rm rel}^{-2/3}. 
\end{equation}

The cooling LF can be $\gamma_{\rm c}$ can be estimated as 
\begin{equation}
    \gamma_{\rm c} = \frac{6 \pi m_{\rm e} c}{\sigma_{\rm T} B^2} \frac{1}{t} =  \left[ \frac{6 \pi m_{\rm e} c}{\sigma_{\rm T}}   \left( \frac{2 \pi m_{\rm e} c}{q_{\rm e}}\right)^{-2/3}  \left( \frac{q_{\rm e}}{\sigma_{\rm T}} \right)^{-4/3} \right] \left( \frac{3p-1}{p-2} \right)^{-4/3} \; \nu_{\rm m}^{-2/3} \; L_{\nu_{\rm m}}^{-4/3} \; E_{\rm rel}^{4/3}\frac{1}{t} 
\end{equation}
while the cooling frequency $\nu_\mathrm{c}$ can be represented as 
\begin{equation}
\begin{split}
 \nu_{\rm c} \equiv \gamma_{\rm c}^2 \nu_{\rm B} = \frac{18 \pi m_{\rm e} q_{\rm e}}{\sigma_{\rm T}^2} \frac{1}{t^2} \frac{1}{B^3} &\ = 9 \left( \frac{p-2}{3p-1} \right)^2 \; \nu_{\rm m}^{-1} \; L_{\nu_{\rm m}}^{-2} \; E_{\rm rel}^2 \; t^{-2} \\
&\  = 18 \left( \frac{p-2}{3p-1} \right)^2 \; E_{\rm rel}^2 \; \left( E_{\rm SN} + E_{\rm rot} \right) \;  R^{-2}\; M_{\rm SNR}^{-1}\; \nu_{\rm m}^{-1} \; L_{ \nu_{\rm m}}^{-2}. \label{freq_cool}
\end{split}
\end{equation}
where in the second line we have replaced $t$ by $t = \frac{R}{v_{\rm SNR}}$ (a relation valid for the coasting phase) and then we have replaced the velocity of SNR by $v_{\rm SNR} = \sqrt{\frac{2 (E_{\rm rot} + E_{\rm SN})}{M_{\rm SNR}}}$. 

The self-absorption $\nu_\mathrm{sa} < \nu_{\rm m}$
can be written as 
\begin{equation}
    \nu_{\rm sa} = \left[ \frac{1}{8 \pi^2 m_{\rm e}}   \left( \frac{2 \pi m_{\rm e} c}{q_{\rm e}}\right)^{-1/3}  \left( \frac{q_{\rm e}}{\sigma_{\rm T}} \right)^{1/3}\right]^{3/5} \left( \frac{3p-1}{p-2} \right)^{1/5} \nu_{\rm m}^{-2/5}\; L_{\nu_{\rm m}}^{4/5}\; E_{\rm rel}^{-1/5}\; R^{-6/5} \label{freq_abs}
\end{equation}

The radius $R$ can be eliminated from equations(\ref{freq_cool})-(\ref{freq_abs}) to give the following relation
\begin{equation}
    \frac{\nu_{\rm c}}{\nu_\mathrm{sa}^{5/3}} = \left[ 144 \pi^2 m_{\rm e} \left( \frac{2 \pi m_{\rm e} c}{q_{\rm e}}\right)^{1/3} \left( \frac{\sigma_{\rm T}}{q_\mathrm{e}} \right)^{1/3} \right] \left( \frac{p-2}{3p-1} \right)^{\frac{7}{3}}   \nu_{\rm m}^{-1/3} L_{\nu_{\rm m}}^{-10/3} E_{\rm rel}^{7/3} \left( E_{\rm SN} + E_{\rm rot} \right) M_{\rm SNR}^{-1}.  
\end{equation}

 Thus,the following relation exists 
\begin{equation}
    \nu_{\rm sa}^{-5/3} \; \nu_{\rm m}^{1/3} \; \nu_{\rm c} \; L_{\nu_{\rm m}}^{10/3} = K\; E_{\rm rel}^{7/3} E_{\rm tot} \; M_{\rm SNR}^{-1} = \frac{K}{2} \; E_{\rm rel}^{7/3} v_{\rm SNR}^2, \label{closure_reln}
\end{equation}
where the constant  $K = 144 \pi^2 m_{\rm e} \left( \frac{2 \pi m_{\rm e} c}{q_{\rm e}}\right)^{1/3} \left( \frac{\sigma_{\rm T}}{q_\mathrm{e}} \right)^{1/3} \left( \frac{p-2}{3p-1} \right)^{{7}/{3}}$ and $E_{\rm SNR} = E_{\rm SN} + E_{\rm rot}$ is the total kinetic energy imparted to the ejecta.

\section{Rotational parameter space of B24}\label{app:B24_Param}

\begin{figure}
\centering
\includegraphics[scale=0.53]{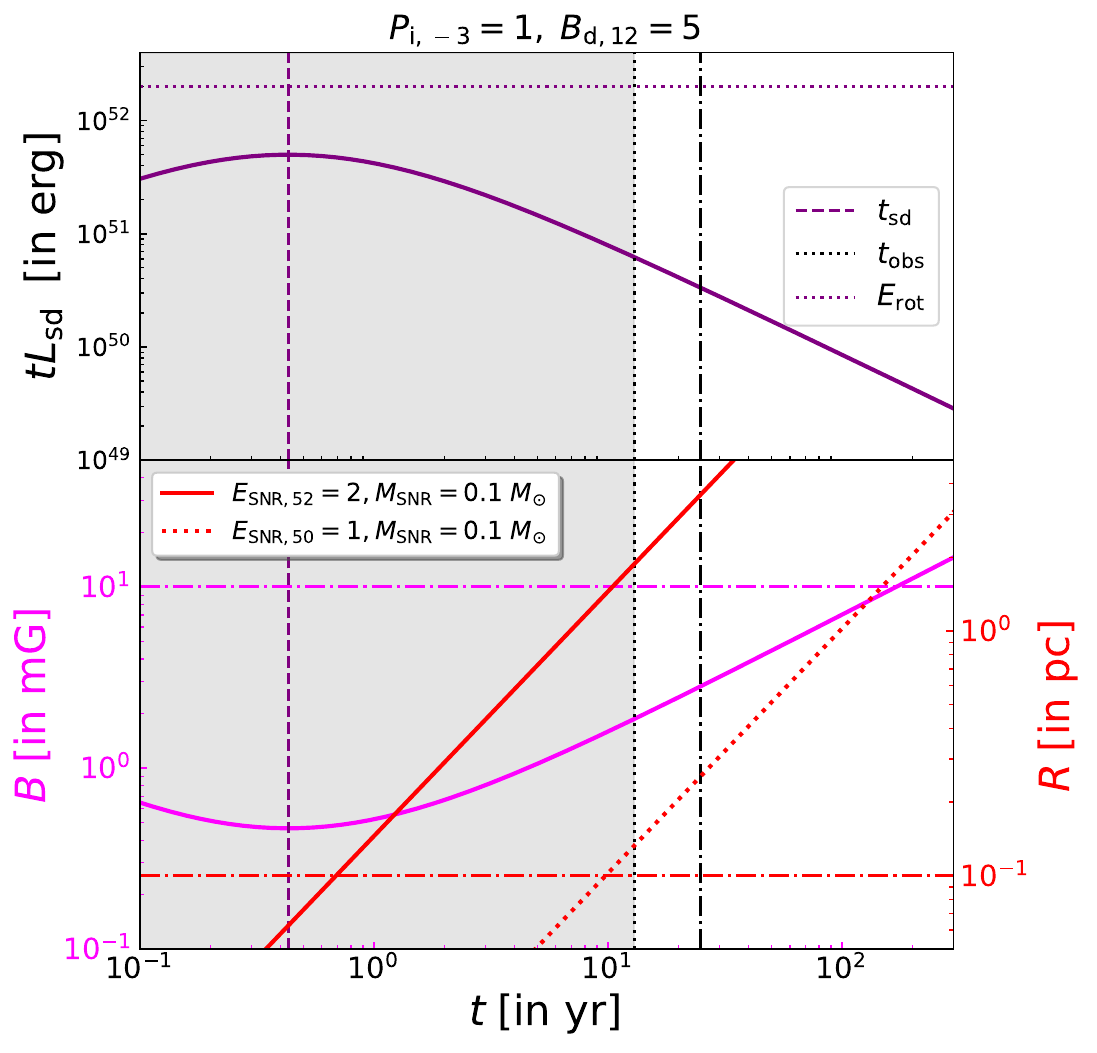}
\caption{Unfeasibility of rotation-powered parameter presented space by B24 for QRS121102 (PRS associated with FRB 121102). In accordance with B24 we assume a neutron star  $P_\mathrm{i} = 1$ ms and a dipolar magnetic field of $B_\mathrm{d} = 5 \times 10^{12}$ gauss in an ultra-stripped supernova with initial explosion energy $E_\mathrm{SN} = 10^{50}$ erg and ejecta mass $M_\mathrm{SNR} = 0.1 M_{\odot}$, and a conversion efficiency of nebular energy to non-thermal energy of electrons to be $\epsilon_\mathrm{e} = 0.99$. The grey shaded region to the left of the black dotted line (the minimum age of QRS121102)  is inaccessible. The suggested age of the nebula $t=25$ years is shown as a vertical dot dashed black line in both panels. For this analysis we also fixed the observables $\nu_{\mathrm m} = 2$ GHz and $L_{\nu_\mathrm{m}} = 2 \times 10^{29}$ erg $s^{-1}$ Hz$^{-1}$. \textit{Top panel:} shows the rotational energy injected in the nebula as a function of time. \textit{Bottom panel:} shows the average nebular magnetic field (in mG) (shown in solid magenta line) on the left and the radius of the nebula (in pc) in red solid line (accounting for SNR expansion boosting due to adiabatic losses) and in red dotted line (without accounting for nebular expansion boost) (see text and \S\ref{rob_conc} for elaborate discussion). \label{fig:B24_param} }
\end{figure}

Figure \ref{fig:B24_param} shows the implication of the parameter space proposed by B24. At the proposed age $t=25$ years (shown as a vertical dot dashed vertical line), neither is the average nebular magnetic field greater than 10 mG violating the findings of \citealt{2022MNRAS.510.4654B} nor does it satisfy the constraint that the radius should be around 0.1 pc as suggested by equipartition analysis (see \S\ref{spec_radio}) and radio scintillation studies by \citealt{2023ApJ...958..185C}. 

\section{Free-Free optical depth}\label{ap:free_free}

The free-free optical depth from supernova (SN) ejecta is given by \citep{Rybicki1979},
\begin{equation}
    \tau_{\rm ff} = 0.018 Z^2 \nu^{-2} T_{\rm ej}^{-3/2} n_{\rm e} n_{\rm ion} g_{\rm ff} R_{\rm ej}
\end{equation}

Assuming $ n_{\rm e} = n_{\rm ion} $, $ Z = 1 $, the expression simplifies to
\begin{equation}
    \tau_{\rm ff} = 0.1 g_{\rm ff} \left(\frac{f_{\rm ion}}{0.1}\right)^2 \nu_{\rm GHz}^{-2} T_4^{-3/2} t_{10}^{-5} (1 + E_{\rm MWN, 51})^{-5/2} \left(\frac{M}{3 M_{\odot}}\right)^{9/2}
\end{equation}
where $T_4 = \frac{T_{\rm ej}}{10^4 \, \text{K}} $. We want to obtain the lower limit on the age of the object when $ \tau_{\rm ff} = 1 $. Inverting the expression,
\begin{equation}
    t_{10} = 0.63 g_{\rm ff}^{1/5} \left(\frac{f_{\rm ion}}{0.1}\right)^{2/5} \nu_{\rm GHz}^{-2/5} T_4^{-3/10} (1 + E_{\rm MWN, 51})^{-1/2} \left(\frac{M}{3 M_{\odot}}\right)^{9/10}
\end{equation}
Assuming $g_{\rm ff} = 1 $, $ T_4 = 1 $, and $ \nu = 0.4 \, \text{GHz}$, the lower limit turns out to be
$t_{10} \sim 1 \quad \text{and} \quad 0.2 \quad \text{for} \quad P_{\rm i} = 10 \, \text{ms} \quad \text{and} \quad 1 \, \; \text{ms}$ respectively.

\section{DM contribution from SNR ejecta}\label{ap:DM_contri}

DM observations associated with FRBs can provide useful constraints. We estimate the upper limit to the ejecta mass by assuming that  the SNR ejecta alone provides the total on-source DM contribution. Following \citealt{Piro2016}, we assume a fraction $f_\mathrm{ion}$ of the ejecta is ionized which translates to a mass estimate as
\begin{equation}
    f_{\rm ion} M_{\rm SNR} \leq 0.3 \; M_{\odot} \; R_{17.3}^2 \; \left( \frac{\mathrm{DM} }{340 \; \text{pc cm$^{-3}$}} \right)         .
\end{equation}
which shows that for the given dispersion measure, a massive ejecta should be accompanied by low ionization fraction. 
\section{Glossary of symbols}

\begin{deluxetable*}{rc}
\tablewidth{0pt}
\tablecaption{Symbols and their meanings \label{tab:glossary}}

\tablehead{
\colhead{Symbol} & \colhead{Meaning}
}
\startdata
$\nu$ & Observation frequency \\
$L_{\nu}$ & PRS spectral luminosity \\ 
$B$ & Nebular magnetic field \\ 
$\sigma$ & Nebular magnetization \\ 
$B_{\mathrm{d}}$ & Equatorial surface magnetic dipole field \\ 
$B_{\mathrm{int}}$ & R.M.S internal magnetic field strength \\
$f_{\mathrm{dip}}$ & Ratio $B_{\mathrm{d}}^2/B_{\mathrm{int}}^2$ \\ 
$\alpha_{\mathrm{B}}$ & Internal field decay index \\
$m$ & Spindown index ($m=2$ for dipole braking) \\ 
$t_{\mathrm{d}}$ & Internal magnetic field decay timescale \\ 
$v_{\mathrm{SNR}}$ & SNR expansion velocity \\
$M_{\mathrm{SNR}}$ & SNR mass \\        
$f_{\mathrm{ion}}$ & SNR ionization fraction \\ 
$E_{\mathrm{SN}}$ & Supernova explosion energy \\
$P_{\mathrm{i}}$ & Initial neutron star spin period \\
$I$ & Neutron star moment of inertia \\
$\Omega_{\mathrm{i}}$ & Initial spin frequency ($= 2\pi/P_{\mathrm{i}}$) \\ 
$E_{\mathrm{rot}}$ & Initial rotational kinetic energy \\
$E_{\mathrm{B,0}}$ & Initial internal magnetic energy \\ 
$L_{\mathrm{B}}$ & Magnetic luminosity $\dot{E}_{\mathrm{B}}$ \\
$\epsilon_{\mathrm{out}}$ & Fraction of $L_{\mathrm{B}}$ converted into baryonic outflow \\
$\epsilon_{\mathrm{e}}$ & Fraction of nebular energy in synchrotron electrons \\
$E_{\mathrm{e}}$ & Energy of non-thermal electrons \\
$E_{\mathrm{SNR}}$ & Total energy injected into SNR ($E_{\mathrm{rot}} + E_{\mathrm{SN}}$) \\
$L_{\mathrm{sd}}$ & Spindown luminosity \\
$t_{\mathrm{sd}}$ & Spindown timescale \\  
$\nu_{\mathrm{m}}$ & Minimum synchrotron frequency \\
$\nu_{\mathrm{c}}$ & Cooling synchrotron frequency \\
$\nu_{\mathrm{sa}}$ & Synchrotron self-absorption frequency \\ 
$t_{\mathrm{ST}}$ & Sedov-Taylor timescale (free-coasting to deceleration) \\
$t_{\tau}$ & SNR optical depth timescale ($\tau_{\mathrm{T}} = 1$) \\
$t_{\mathrm{ff}}$ & Free-free absorption timescale ($\tau_{\mathrm{ff}} = 1$) \\ 
$t_{\mathrm{switch}}$ & Timescale when nebular energy transitions from rotation-dominated to internal magnetic field-dominated \\ 
$p$ & Non-thermal electron power-law index \\ 
$\gamma_{\mathrm{m}}$ & Minimal Lorentz factor of non-thermal electrons \\
\enddata
\tablecomments{This glossary provides definitions for all key symbols used in the analysis. 
Subscripts denote specific physical contexts (e.g., $\mathrm{SNR}$ for supernova remnant quantities).}
\end{deluxetable*}

\clearpage       



\end{document}